\documentclass{aa}
\usepackage{txfonts}
\usepackage{natbib}
\bibpunct{(}{)}{;}{a}{}{,}    
\usepackage{graphicx}
\usepackage{subfig}
\usepackage[breaklinks=true, colorlinks=true, urlcolor=magenta, citecolor=blue, linkcolor=blue]{hyperref}

\begin{document}

\title{The \textit{Fermi}-LAT view of the changing-look blazar OQ 334}

\author{S.S. Ren
          \inst{1}
          \and
            R.X. Zhou\inst{2}
          \and
            Y.G. Zheng\inst{1,\star{}}
          \and S.J. Kang\inst{3,\star{}}
          \and Q. Wu\inst{4}
          }
\offprints{Y.G. Zheng}
   \institute{Department of Physics, Yunnan Normal University, Kunming, 650092, China\\
              \email{ynzyg@ynu.edu.cn}
         \and
            Institute of Space Sciences, Shandong University, Weihai, Shandong, 264209, People's Republic of China
         \and
            School of Physics and Electrical Engineering, Liupanshui Normal University, Liupanshui, Guizhou 553004,China\\
            \email{kangshiju@alumni.hust.edu.cn}
         \and
            Department of Astronomy, School of Physics, Huazhong University of Science and Technology, Wuhan, Hubei, 430074, China.
             }

\abstract
   { Unusually, there are still certain characteristics of the changing-look (CL) active galactic nuclei (AGNs) that remain undetected. Consequently, the trigger mechanism behind the CL phenomenon observed in partial AGNs remains unknown.}
   { We explore the light curve and spectral energy distribution (SED) of the CL blazar OQ 334 as obtained by \textit{Fermi}-LAT.}
   { By examining the variability of the equivalent width (EW), we categorise the \textit{Fermi}-LAT light curves of OQ 334 during the epoch of MJD 54628-58677 into seven distinct epochs, including the flat spectrum radio quasar (FSRQ) state, the transition state, and the BL Lac state. We obtained both a \textit{Fermi}-LAT SED and a multi-wavelength SED for each of these distinct epochs.}
   { The source exhibits a transformation from a quiescent state to a highly active state, as evidenced by the variability of the EW. The multi-wavelength SEDs display a prominent external Compton characteristic, even though the \textit{Fermi}-LAT SED reveals both a FSRQ and a BL Lac state across the seven different epochs. To gain further insights, we employed a leptonic model that takes into account the soft photon fields originating from both synchrotron radiation and the external environment. By simulating the multi-wavelength SEDs for each epoch, we uncover the following results. Firstly, the energy density of the external photon fields evolves in an oscillatory manner over the seven different epochs. Also, the energy density of the external photon fields in the BL Lac state is lower than that in the FSRQ state. }
   {These findings suggest that the CL blazar represents a unique phase in the blazar sequence. Considering that the energy density of the external photon fields is proportional to the accretion rate, we propose that evidence for the interconversion of advection-dominated accretion flow (ADAF) discs and standard Shakura-Sunyaev discs (SSDs), --- as variations in accretion modes in the CL blazar --- can be obtained through observations by \textit{Fermi}-LAT.}


\authorrunning{S.S. Ren et~al.}
\maketitle


\section{Introduction}
\label{sec:intro}
Blazars, a subclass of active galactic nuclei (AGNs), exhibit a peculiar characteristic where the angle between the jet direction of the blazar and the line of sight is extremely small. The multi-wavelength spectral energy distribution (SED) serves as a representation of the energy-flux density of blazars across different frequencies. The SED prominently displays two distinct humps. The first hump, known as the low-energy hump, is typically observed within the optical to X-ray wavelength range. The second hump, referred to as the high-energy hump, is usually observed between X-ray and $\gamma$-ray. The generation of the lower energy hump in blazars is widely accepted as being a result of the relativistic electron synchrotron (Syn) radiation occurring within the jet \citep{Urry98}. However, the mechanism behind the generation of the high-energy hump is still a subject of controversy, with two main explanations being the leptonic model and the hadronic model. In the leptonic model, the second hump is typically attributed to inverse Compton (IC) scattering. Specifically, when the low photon involved in the IC \citep{Maraschi92} originates from low-energy Syn radiation, the high-energy hump is generated through synchrotron self-Compton (SSC) radiation \citep{Marscher85}. Additionally, when the low-energy photons for the IC come from outside of the jet, the high-energy hump is generated through external Compton (EC) radiation. The external photon field has four main sources: (1) photons emitted from the accretion disc \citep{Dermer92, Sikora94}; (2) photons originating from the broad-line region (BLR) \citep{Sikora94, Blandford95,1996MNRAS.280...67G,1998MNRAS.294..439B};
(3) photons derived from infrared radiation emitted by the dust torus (DT) outside the jet zone, and microwave background radiation \citep{1995ARA&A..33..163W}.

The SSC model has been widely employed to fit the multi-wavelength SED of BL Lacertaes (BL Lacs) \citep{2004ApJ...601..151K,2012ApJ...752..157Z}, while a combination of SSC and EC is often used to describe flat-spectrum radio quasars (FSRQs) more accurately \citep{2002ApJ...581..127B,2011MNRAS.414.2674G,2014MNRAS.439.2933Y,2014ApJ...788..104Z,2019NewAR..8701541H}.
Blazars can also be categorised based on the peak frequency of their low-energy Syn radiation: low-synchrotron-peaked objects (LSP, $v_{peak}^{syn} < 10^{14}$ Hz), intermediate-synchrotron-peaked objects (ISP, $10^{14}$ Hz < $v_{peak}^{syn}$ < $10^{15}$ Hz), and high-synchrotron-peaked objects (HSP, $v_{peak}^{syn} > 10^{15} $Hz) \citep{2010ApJ...716...30A}.
Another categorisation of blazars is based on the equivalent width (EW) of their emission lines \citep{1991ApJS...76..813S,1995PASP..107..803U}.
BL Lacs have weak or no emission lines if EW < 5 {\AA}, while FSRQs exhibit strong emission lines, with EW > 5 {\AA}.

OQ 334, also known by other names such as 
QSO B1420+326 \citep{2022icrc.confE.775D}, B2 1420+32 \citep{2021ApJ...913..146M}, and FSRQ QSO B1420+326 \citep{2021A&A...647A.163M}, is classified as a FSRQ and falls under the category of blazars in the 4FGL catalogue \citep{2020ApJS..247...33A}.
It has a redshift value of 0.6819 \citep{2010MNRAS.405.2302H}.
Since the launch of \textit{Fermi Gamma-Ray Space Telescope Large Area Telescope} (\textit{Fermi}-LAT) in 2008, OQ 334 has been observed in $\gamma$-ray at relatively low brightness \citep{2021MNRAS.502.5245P}.
However, in 2018, a $\gamma$-ray flare was detected \citep{2018ATel12277....1C}.
Two years later, another flare was observed by \textit{Fermi}-LAT \citep{2020ATel13382....1C,2022ATel15793....1K}.
\cite{2021ApJ...913..146M} found significant evidence of the changing nature of OQ 334 (also known as B2 1420+32).
Before January 4, 2018, this object was observed as a FSRQ, but between January 4 and 6, 2018, it transitioned into a BL Lac state, before changing back into a FSRQ state again 96 days later. This phenomenon of transitioning between different types is referred to as the changing look (CL) phenomenon. OQ 334 has been identified as a CL blazar.

The category of CL AGNs is defined when certain conditions are met.
One condition involves the observation of the phenomenon of broad emission lines turning on or turning off over a timescale of several years in the optical spectrum of certain AGNs \citep{2014ApJ...788...48S,2015ApJ...800..144L,2016MNRAS.455.1691R,2016A&A...593L...8M,2016MNRAS.461.1927P,2016ApJ...826..188R,
2018ApJ...862..109Y,2019MNRAS.486..123R,2020MNRAS.491.4925G,2020ApJ...901....1W,2020A&A...638A..91K}.
Another condition pertains to X-ray observations, wherein the state of some AGNs transitions between the Compton-thick and the Compton-thin states on timescales ranging from hours to months \citep{2003MNRAS.342..422M,2015ApJ...800..144L,2016ApJ...820....5R,2017ApJ...840...11L}.
Additionally, there have been instances where certain Seyfert galaxies have transformed into different types of Seyfert galaxies within a few years, while some of them eventually revert back to their original state \citep{1992AJ....104.2072T,1993ApJ...410L..11S,1999ApJ...519L.123A,2014ApJ...796..134D}.
The identification of an increasing number of  CL AGNs is the result of extensive searches within the vast expanse of AGNs \citep{2016AJ....151...32A,2018A&A...619A.168K,2020ApJ...890L..29A}.
The information we have on these phenomena holds great significance in our understanding of various processes, such as black hole accretion physics, the co-evolution of AGNs and galaxies, and the role of AGNs in their host galaxies.

The phenomenon of CL has attracted the attention of numerous authors aiming to understand the underlying causes.
Two prevailing arguments have emerged to explain this phenomenon.
Firstly, it has been proposed that CL occurs due to variable obscuration, wherein the obscuring material moves in or out along the line of sight of the observer.
This movement leads to the appearance or disappearance of the broad emission lines \citep{2013MNRAS.436.1615M,2014MNRAS.443.2862A,2015ApJ...815...55R,2018MNRAS.481.2470T}.
Alternatively, variation in the accretion rate of the black hole has been proposed to contribute to the CL phenomenon \citep{2021ApJ...909...18F}.
For instance, a dramatic decrease in the accretion rate results in a significant reduction in high-energy radiation from the accretion disc, which is then not sufficient to produce broad emission lines.
Conversely, an enhancement in the supermassive black hole (SMBH) accretion rate can give rise to the generation of broad emission lines.
AGNs associated with the former interpretation are often referred to as changing-obscuration AGNs (CO-AGNs), while those related to the latter interpretation are termed as changing-state AGNs (CS-AGNs) \citep{2023NatAs...7.1282R}.

\cite{2021ApJ...913..146M} propose that the CL phenomenon observed in OQ 334 is a result of a significant change in the continuous flux of the jet, which dilutes the spectral broad emission line and allows for transitions between FSRQ and BL Lac states \citep{2023MNRAS.525.3201K}.
Changing-look blazars may be due to jet variability or accretion disc variation.
The aim of this work is to investigate the underlying mechanisms responsible for the continuous radiation of the jet, with a specific focus on $\gamma$-ray emissions, and to try to explain the mechanism that triggers the CL phenomenon in OQ 334.

This paper is organized as follows. Section 2 provides details on the data sources and data compilation.
In Sect. 3, we present the model and explain the parameter adjustments made during the analysis. Section 4 applies the developed model to OQ 334, presenting  the obtained results. Section 5 discusses the observed phenomena and their implications. Finally, in Sect. 6, we summarise our findings. Throughout this work, we consider the following cosmological constants:
$H_{0}=67.8$ km s$^{-1}$ Mpc$^{-1}$, $\Omega_{\rm M}=0.308$, $\Omega_{\rm r}=0$, and $\Omega_{\Lambda}=0.692$ \citep{2016A&A...594A..13P}.


\section{Data reduction and analysis}   
\label{sec:data}
\subsection{ $\gamma$-ray data}   

We performed an analysis of $\gamma$-ray data provided by \textit{Fermi}-LAT covering the energy range of 0.1-300 GeV from August 11, 2008, to July 12, 2019.
To handle the large volume of irregular and 
discontinuous Fermi observation data, we employed the \textit{Fermitools} toolkit. The photon file and spacecraft file (at the end of \textit{$\ast\ast\ast$.fits}) observed by \textit{Fermi}-LAT were processed into the required format (ending in \textit{$\ast\ast\ast$.csv}), which is needed to use in the prompt of the Likelihood Analysis With Python website\footnote{\scriptsize{\url{https://fermi.gsfc.nasa.gov/ssc/data/analysis/scitools/python_tutorial.html}}}.
The \textit{gtselect} package was used to select and extract the event data file, and to select the coordinates of OQ 334 as the center at the same epoch.
A region of interest (ROI) radius of 10 deg was chosen to determine the source area, and photons with energies ranging from 100 MeV to 0.3 TeV were selected.
Additionally, to mitigate the earth's albedo contamination, we implemented a zenith angle cut of less than ${90^\circ }$.
Subsequently, the \textit{gtmktime} package was used to select spacecraft data file variables and create good time intervals (GTIs).
The \textit{gtbin} package was employed to generate an ROI count map for subsequent spectrum generation.
The \textit{gtltcube} and \textit{gtexpmap} packages were used to produce the exposure map.
The galactic interstellar emission model\footnote{\scriptsize{\url{https://fermi.gsfc.nasa.gov/ssc/data/access/lat/BackgroundModels.html}}} (\textit{gll\_iem\_v07. fits}) and isotropic spectral template (\textit{iso\_P8R3\_SOURCE\_V3\_v1.txt}) were used to create an XML format file using the generated \textit{4FGLxml.py} script \citep{2020ApJS..247...33A}; see for example \citealt{2021ApJ...916...93Z}.
The diffuse sources contained in the source model XML file were created using \textit{gtdiffrsp}, and the analysis was performed using the \textit{gtlike} task.
To quantify the $\gamma$-ray analysis, we employed the maximum likelihood test statistic $TS=2\Delta log(L)$, where L represents the ratio of the likelihood values between the model with $\gamma$-ray point objects and the model without $\gamma$-ray objects \citep{1996ApJ...461..396M,2022MNRAS.511..938Z,2020ApJS..247...33A}.
Combining these steps with the open-source Python software package \textit{Fermipy} \citep{2017ICRC...35..824W} allowed us to calculate the $\gamma$-ray photon spectrum of OQ 334 using \textit{Fermi}-LAT data.


\subsection{\textit{X-ray} data}

The \textit{Swift} XRT is an X-ray imaging spectrograph known for its notable advantages \citep{2005SSRv..120..165B}.
Firstly, it exhibits sensitivity in the 0.2-10 KeV energy band.
Secondly, it offers two timing modes, namely the photon counting (PC) mode and the windowed timing (WT) mode, enabling accurate photometric measurement and the generation of photometric curves with a time resolution of at least 10 milliseconds.
To select X-ray data that are quasi-simultaneous with gamma rays, we retrieved data from the High Energy Astrophysics Science Archive Research Center (HEASARC) website\footnote{\scriptsize{\url{https://heasarc.gsfc.nasa.gov/db-perl/W3Browse/w3browse.pl}}} spanning from April 12, 2005, to July 13, 2019 (see Table~\ref{tab:01}).
The data were analysed according to standard threads\footnote{\scriptsize{\url{https://www.swift.ac.uk/analysis/xrt/index.php}}} for level I data.
The \textit{Swift} XRT data were restored using \textit{HEASoft 6.26.1}.
The source area file was selected from a central circle with a radius of approximately $ \sim20 $ pixels (47 arcsecs), while the background area file was chosen from a ring with internal and external radius of approximately $ \sim51 $ pixels (120 arcsecs) and $\sim85 $ pixels (200 arcsecs), respectively.
The level II data from both WT mode and PC mode were used to generate the light curves and spectra using the \textit{xselect} tool. The spectra from WT mode and PC mode were binned using \textit{grppha}, with a minimum requirement of 20 and 10 photons per bin, respectively.
To restore the X-ray spectra, specific response matrix files (\textit{swxwt0to2s6\_20131212v015.rmf}, \textit{swxpc0to12s6\_20130101v014.rmf}) were employed for WT mode and PC mode, respectively.
Standard auxiliary response files were created using \textit{xrtmkarf}.
We used the \textit{xspec 12.10.1} software package to fit the grouped optical spectrum, and the redshift power law (zPL) model to fit the energy spectrum.
\subsection{Optical/ultraviolet, and other data}    

The Ultraviolet Optical Telescope (UVOT, \citealt{2005SSRv..120...95R}) on the Swift satellite was employed for the analysis of all the observed data within the MJD 54628-58677 epoch interval.
The UVOT instrument uses six filters to capture optical and ultraviolet (UV) data across various wavelengths: V (500 - 600 nm), B (380 - 500 nm), U (300 - 400 nm), UVW1 (220 - 400 nm), UVM2 (200 - 280 nm), and UVW2 (180 - 260 nm).
Following the recommended threads\footnote{\scriptsize{\url{ https://www.swift.ac.uk/analysis/uvot/index.php}}}, the source and the background files were extracted from circles with radii of ${5^\circ }$ and ${20^\circ }$, respectively, using \textit{HEASoft 6.26.1}.
Subsequently, the flux values were adjusted for reddening using a value of E (B-V)=0.093 mag.
The magnitudes corresponding to the V, B, U, UVW1, UVM2, and UVW2 bands were corrected by 0.173, 0.229, 0.275, 0.394, 0.457, and 0.481 mag \citep{2015MNRAS.454..353R}), respectively.

To compile the multi-frequency flux observation data of OQ 334, we gathered the available data from the ASI Space Science Data Center (SSDC\footnote{\scriptsize{\url{ https://tools.ssdc.asi.it/SED/}}}) and calculated their averages based on the methodology proposed by \cite{2021ApJ...915...59Z}.
Additionally, the SSDC website was consulted to obtain further potentially useful data for analysis.


\section{The model}\label{sec:model} 

\subsection{\textit{Fermi}-LAT SED}   

We conducted a standard \textit{Fermi}-LAT likelihood analysis on the initial \textit{Fermi}-LAT observations of OQ 334.
The spectrum of OQ 334 was modelled using a power-law spectrum \citep{2002PASJ...54..533F,2013PASJ...65...25F,2010ApJ...722..520A}) to obtain the relation between $\gamma$-ray $log\nu$ and $log\nu f_{\nu}$:

\begin{equation}
\frac{d N}{d E}=N_{0}(\frac{E}{E_{0}})^{-\Gamma_{\gamma}}
\label{eq:1}
\end{equation}
where $\Gamma_{\gamma}$ represents the photon spectrum index, and $N_{0}$ denotes the initial flux.

\subsection{Multi-wavelength SED}   

In this work, we employed the traditional Syn+ SSC+ EC model \citep{1999ApJ...515..140S,2011ApJ...735..108C,2016MNRAS.461.1862K} to fit the SED of OQ 334.
The model assumes the presence of a spherical radiation region with a radius $R$, containing extremely relativistic electrons and a disordered, uniform magnetic field B.
It is further assumed that the homogeneous sphere moves outward relative to the jet with a velocity of $ \nu = c\beta $ along the jet direction, where $c$ represents the speed of light, $\beta =\sqrt{1-1/{\Gamma}^{2}} $ , and $\Gamma$ is the bulk Lorentz factor.
For relativistic jets with narrow observation angles $\theta \leq 1/\Gamma $, the Doppler factor is $\delta =[\Gamma(1-\beta \cos \theta)]^{-1} \approx \Gamma $.
The electron spectrum is characterised by a log-parabolic distribution generated through random acceleration, as described by Eq. (\ref{eq:3}) \citep{2004A&A...413..489M,2004A&A...422..103M,2009A&A...501..879T,2011ApJ...739...66T,2014ApJ...788..179C}:

\begin{large}
\begin{equation}
    N(\gamma) = {K_0}(\frac{\gamma}{{\gamma_0}})^{-s-b~log(\frac{\gamma}{{\gamma_0}})}
           \qquad    \gamma_{min} < \gamma < \gamma_{max}
    \label{eq:3}    
\end{equation}
\end{large}

where, $s$ denotes the spectral index, $b$ represents the spectral curvature, $\gamma_0 $ is the initial Lorentz factor, and $ K_0 $ is the normalisation constant of the logarithmic parabolic electron spectrum.

The energy-flux density of EC radiation is calculated using the following formula in the observer coordinate system:

\begin{large}
\begin{equation}
    \begin{array}{l}
    {\nu F_{{\rm{\nu}}}} = \frac{{3c\sigma _{T}\delta _{}^4\epsilon_{{\rm{\gamma}}}^{\prime2} }}{{{16\pi d_{L}^{2}}}}\int_{0}^{\infty} {\frac{{{u_{ext}^{\prime}(\epsilon^\prime)}}}{{{\epsilon^{\prime 2}} }}}d{\epsilon ^\prime } \\  \;\;\;\;\;\;\;\;\;\;\;\;\;\;\;\;
    \times \int_{\gamma_{min}^{\prime}}^{\gamma_{max}^{\prime}} {\frac{{N_{e}^{\prime}({\gamma ^\prime })}}{{{\gamma ^\prime }^2}}F_{c}(q^\prime ,\Gamma _{{\rm{e}}}^\prime )} d{\gamma ^\prime }
    \end{array}
    \label{Eq:4}
\end{equation}
\end{large}

In Eq. (\ref{Eq:4}), $\epsilon_{\gamma}^{\prime}=\frac{h\nu^{\prime}}{m_{e}c^{2}}$ represents the dimensionless energy of the scattered photon in the co-moving coordinate system, $\nu^{\prime}$ corresponds to the frequency of the scattered photon,
$u_{ext}^{\prime}$ denotes the energy density of the external photon field (Count $u_{ext}^{\prime}$ as $U_{ext}$) \citep{2012PASJ...64...80Y,2019ApJ...873....7Z,2020MNRAS.499.1188Z}.
$F_{c}(q^\prime,\Gamma _{{\rm{e}}}^\prime )$ denotes the Compton scattering kernel, as described in Eq. (\ref{Eq:5}) \citep{2008ApJ...686..181F,2009ApJ...692...32D}), and $d_{L}$ represents the luminosity distance \citep{2007NJPh....9..445C,2015IJMPA..3045020F}), as described $d_{L} = (1+z)\frac{c}{H_{0}} \int_{1}^{1+z}\frac{1}{\sqrt{\Omega_{M}x^{3}+1-\Omega_{M}}} d{x}$.
Both the SSC and EC models in the present study take into account the Klein-Nishina (KN) effects.

\begin{large}
\begin{equation}
    \begin{array}{l}
    F_{c}(q^\prime ,\Gamma _{{\rm{e}}}^\prime ) = [2q^\prime lnq^\prime + (1+2q^\prime)(1-q^\prime) \\  \;\;\;\;\;\;\;\;\;\;\;\;\;\;\;\;
    +\frac{1}{2}\frac{(\Gamma_{e}^{\prime}q^\prime)^2}{(1+\Gamma_{e}^{\prime}q^\prime)}(1-q^\prime)]
    H(q^\prime;\frac{1}{4\gamma^{\prime 2}},1)
    \end{array}
    \label{Eq:5}
\end{equation}
\end{large}
where $q^\prime \equiv \frac{\epsilon_{s}^{\prime}/\gamma^\prime}{\Gamma_{e}^{\prime}(1-\epsilon_{s}^{\prime}/\gamma^\prime)}$, and $\Gamma_{e}^{\prime}=4\epsilon^{\prime}\gamma^\prime$.

We propose that the seed photons of the IC process originate from two sources: photons generated by Syn radiation within the jet and the photons emitted by the BLR and DT located outside the jet's radiation region.
To investigate the variations in the energy density of the external photon field during the CL phenomenon and the source of seed photons from the external photon field, we treat the radiation region location $r_{dis}$ and the energy density of the external photon field $U_{ext}$ as free parameters. We use $r_{dis}$ to denote the distance of the emitting region from the central source.

Our model encompasses a total of 11 parameters: $s$, $b$, $\gamma_{min}$, $\gamma_{max}$, $\gamma_{0}$, $K_{0}$, $\delta$, B, R, $r_{dis}$, and $U_{ext}$.
The parameters $\gamma_{min}$ and $\gamma_{max}$ were found to have no significant impact on the fitting results and so we fixed them to reduce the number of model fitting parameters. We adopted a typical value of $\gamma_{min}$ =50 and $\gamma_{max}=1\times 10^{8}$ (e.g., \citealt{2014ApJS..215....5K,2014ApJ...788..104Z,2016MNRAS.461.1862K,2021MNRAS.502.5875K}).
The radius of the radiation region is estimated using the typical minimum variability timescale $\Delta t_{\mathrm{var}}=1$ day (e.g., \citealt{1998MNRAS.301..451G,2008ApJ...677..906F,2012ApJ...752..157Z}) , which corresponds to $R=c\delta\Delta t_{\mathrm{var}}/(1+z)$ (e.g., \citealt{2009ApJ...697..934A,2013MNRAS.436.2170C}).
The other parameters, namely $s$, $b$, $\gamma_{0}$, $K_{0}$, $\delta$, $r_{dis}$, and B, were kept as free variables in our fitting.
We evaluated the chi-square value ${\chi ^2} = 1/(N - {\rm{dof}})\sum\limits_{i = 1}^N {{{\left( {{{\hat y}_i} - {y_i}/{\sigma _i}} \right)}^2}} $, where ${\rm{dof}}$ corresponds to the number of free parameters; $N$ represents the number of observation data points that participate in the chi-square value calculation; ${{{\hat y}_i}}$ denotes the expected value predicted by the model; ${{y_i}}$ represents the observed data; and ${{\sigma _i}}$ signifies the standard deviation of the data point \citep{2014A&A...567A.135A,2020sdmm.book.....I}.


\section{Result} 
\subsection{\textit{Fermi}-LAT light curve and SED}

\begin{figure}
\centering
\includegraphics[width=0.5\textwidth]{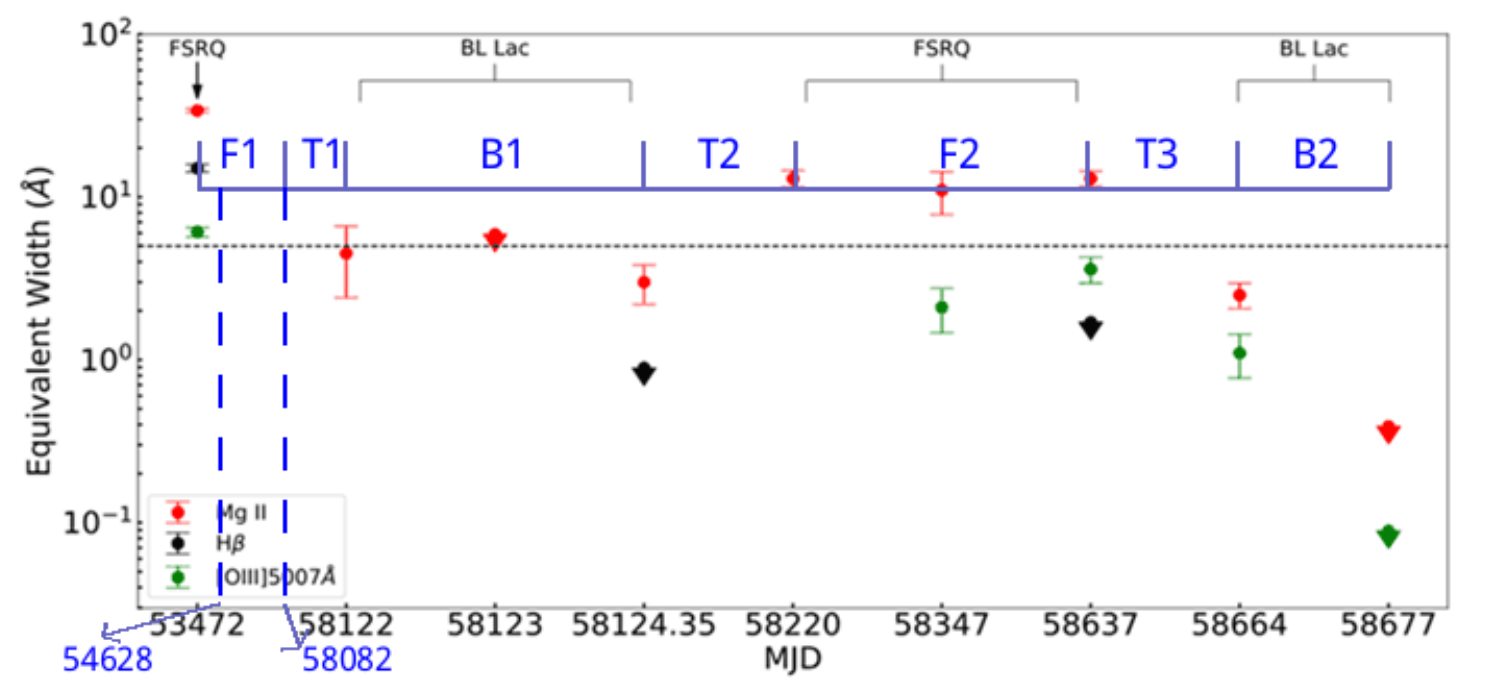}
\caption{CL event of the blazar OQ 334 at MJD 54628-58677 \citep{2021ApJ...913..146M}. According to the FSRQ EW of greater than 5 {\AA} and the BL Lac EW,which is less than 5 {\AA}, the entire CL epoch (MJD 54628-58677, ${F_1}$-${B_2}$) is subdivided into an FSRQ state (${F_1}$, ${F_2}$), a transition state (${T_1}$, ${T_2}$, ${T_3}$), and a BL Lac state (${B_1}$, ${B_2}$).}
\label{Fig01}
\end{figure}

\cite{2021ApJ...913..146M} analysed the variation of the EW of Mg \uppercase\expandafter{\romannumeral2}, H$\beta$, and [O\uppercase\expandafter{\romannumeral3}]5007 {\AA} in the context of the CL phenomenon in OQ 334, demonstrating the existence of oscillatory variations (see Fig.~\ref{Fig01}).
Based on the criterion of whether the EW of the Mg \uppercase\expandafter{\romannumeral2} line exceeds 5 {\AA}, the CL phenomenon of OQ 334 within the MJD 54628-58677 epoch can be categorised into seven epochs: ${F_1}$, ${T_1}$, ${B_1}$, ${T_2}$, ${F_2}$, ${T_3}$, and ${B_2}$ (refer to Table~\ref{tab:01} and Fig.~\ref{Fig01}), with MJD 54628-58677 denoted as ${F_1}$-${B_2}$.
The division of these specific epochs is as follows: an epoch where the Mg \uppercase\expandafter{\romannumeral2} EW (represented by a red dot) exceeds 5 {\AA} is classified as an FSRQ state; an epoch where the Mg \uppercase\expandafter{\romannumeral2} EW falls below 5 {\AA} is classified as a BL Lac state; and an epoch with no red dot is classified as a transitional state between FSRQ and BL Lac.
Notably, the ${F_1}$ epoch is selected as MJD 54628-58082 instead of MJD 53472-58082 due to the use of the $\gamma$-ray data obtained through \textit{Fermi}-LAT, which was launched on June 11, 2008 (MJD 54628).
Additionally, as ${T_1}$ is the same as ${T_3}$, given that OQ 334 transitioned from the FSRQ to the BL Lac state, and the  ${T_3}$ epoch lasted for 27 days, it is assumed that the ${T_1}$ epoch spanned 38 days, resulting in the range MJD 58083-58121 for ${T_1}$.

\begin{figure*}
	\centering
		\includegraphics[width=1\linewidth]{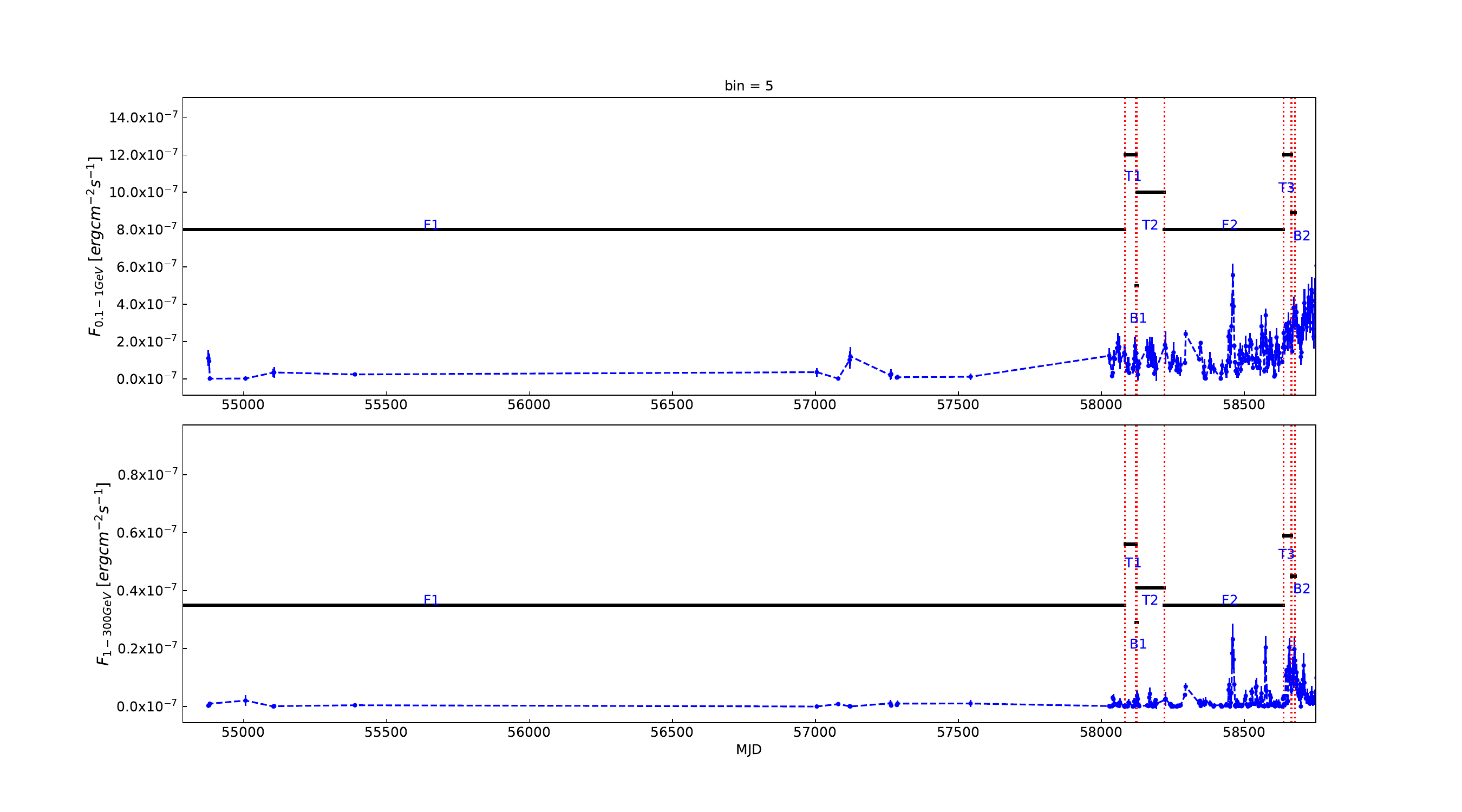}%
  \centering
\caption{\textit{Fermi}-LAT at MJD 54628-58677 (bin=5 days). The flux in the 0.1-1GeV band is indicated by $F_{0.1-1GeV}$ (top panel). Flux in the 1-300 GeV band is expressed as $F_{1-300GeV}$ (bottom panel). The range of the light curve corresponding to ${F_1}$, ${T_1}$, ${B_1}$, ${T_2}$, ${F_2}$, ${T_3}$, and ${B_2}$ of the CL phenomenon is marked by black segments. The data points below certain TS $\leq$ 15 are removed.} 
	\label{Fig02}
\end{figure*}

We analysed the \textit{Fermi}-LAT light curve for the epoch MJD 54628-58677, as depicted in Fig.~\ref{Fig02}.
Each point in the light curve possesses a TS value of greater than or equal to 15.
These authors also analysed the light curves in the 0.1-1 GeV range (low-energy band, denoted $F_{0.1-1GeV}$) and the 1-300 GeV range (high-energy band, $F_{1-300GeV}$) of $\gamma$-rays during the CL phenomenon.
We also calculated the value of fractional flux variability ($F_{var}$; \citealt{1997ApJS..110....9R,2002ApJ...568..610E,2003MNRAS.345.1271V,2014ApJ...783...83L,2018ApJ...866...16P}).
We obtain $F_{var}$=$0.690\pm 0.040$ in the 0.1-1 GeV energy bands, and $F_{var}$=$0.872\pm0.191$ in the 1-300 GeV energy bands.

\begin{figure*}
	\centering
		\includegraphics[width=1\linewidth]{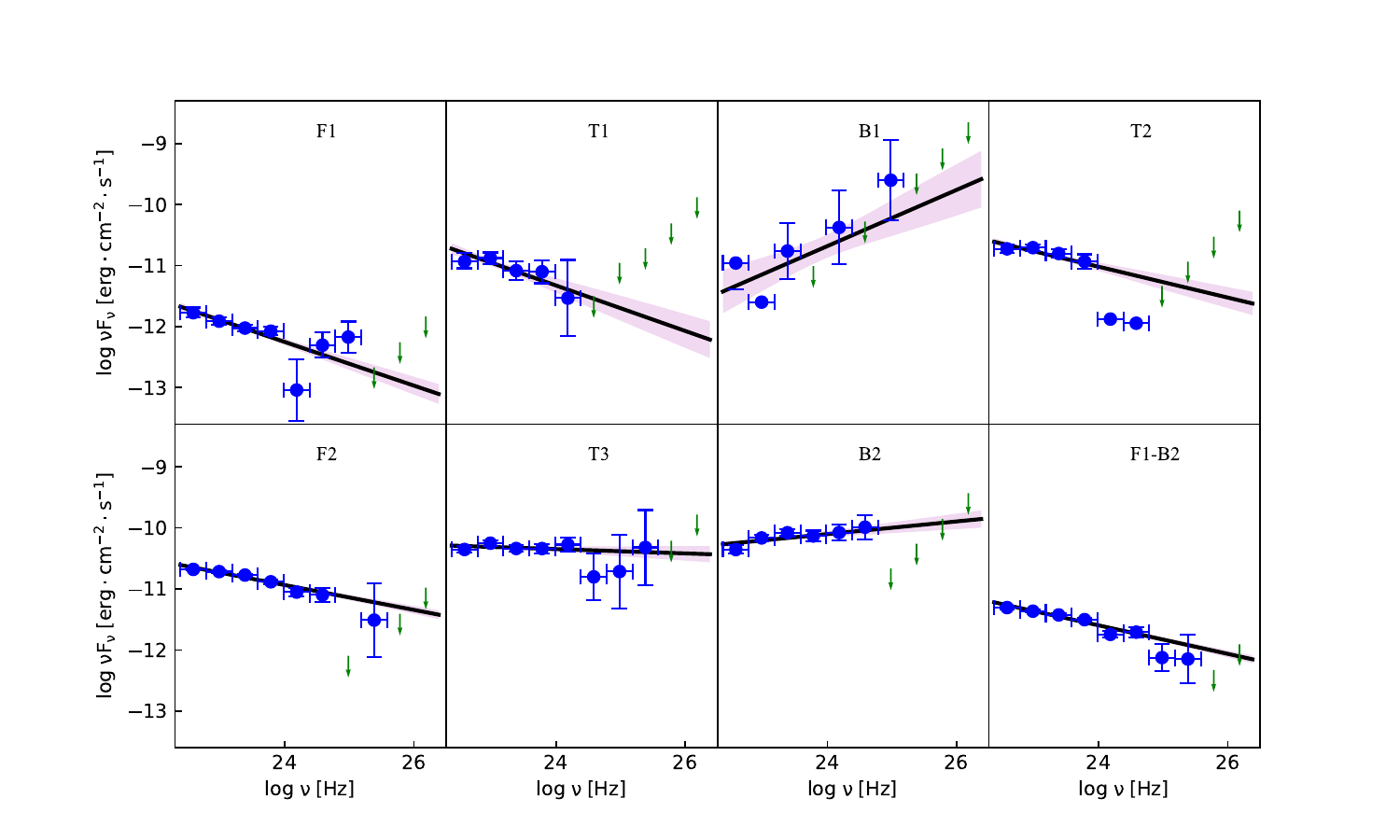}
		\centering
	\caption{\textit{Fermi}-LAT SED for the CL phenomenon. The solid black line is the best fit for the likelihood analysis over the entire energy range. The blue dots are $\gamma$-ray data points, and the green downward arrows indicate the upper energy limit. The purple area is the 1$\sigma$ confidence band. The `letters + numbers' in each subgraph correspond to the status in Table~\ref{tab:01}.} 
	\label{Fig03}          
\end{figure*}

The \textit{Fermi}-LAT SED corresponding to epochs ${F_1}$, ${T_1}$, ${B_1}$, ${T_2}$, ${F_2}$, ${T_3}$, ${B_2}$, and ${F_1}$-${B_2}$ (refer to Fig.~\ref{Fig03})  was fitted using \textit{Fermipy}, resulting in the determination of the photon spectral index $\Gamma_{\gamma}$ for each epoch (see Table~\ref{tab:01}).
We note that the photon spectral index in the transition states (${T_1}$, ${T_2}$, ${T_3}$) and the FSRQ states (${F_1}$, ${F_2}$) was significantly higher than that in the BL Lac states (${B_1}$, ${B_2}$) (see Table~\ref{tab:01}).
Consequently, a correlation between the type transformation and the change in photon spectral index can be inferred.
The photon spectral index of ${F_1}$-${B_2}$ is closer to that of ${F_1}$ and ${F_2}$.
\begin{table*}
	\centering
    \renewcommand\arraystretch{2}
    \tabcolsep=0.2cm
	\caption{\textit{Fermi}-LAT and Swift observation times and \textit{Fermi}-LAT SED results in CL.}
	\label{tab:01}
	\begin{tabular}{cccccccc} 
		\hline \hline
        \makebox[0.01\textwidth][c]{Epoch}          &  \makebox[0.01\textwidth][c]{State}         & \makebox[0.01\textwidth][c]{$\gamma$-ray} & $\gamma$-ray   & $\Gamma_{\gamma}$ & X-ray      & Optical-UV \\
         ...                                        &  ...                                        & MJD                                       & year/month/day  &...       & year/month/day & MJD\\
         (1)                                        &  (2)                                        & (3)                                       & (4)            &(5)       & (6)        & (7)  \\
		\hline
        ${F_1}$           & FSRQ                        &54628-58082 & 2008/6/11-2017/11/25   &2.358$\pm$0.064  & 2016/8/31$\ast$     & $\star$\\
		${T_1}$           & FSRQ$\longrightarrow$BL Lac &58083-58121 & 2017/11/26-2018/1/3    &2.372$\pm$0.129  & $\ast$              & $\ast$ \\
		${B_1}$           & BL Lac                      &58122-58124 & 2018/1/4-2018/1/6      &1.540$\pm$0.322  & 2018/1/20           & $\ast$ \\
		${T_2}$           &BL Lac$\longrightarrow$FSRQ  &58125-58220 & 2018/1/7-2018/4/12     &2.253$\pm$0.072  & 2018/2/22           & 58171.9\\
        ${F_2}$           & FSRQ                        &58221-58637 & 2018/4/13-2019/6/3     &2.204$\pm$0.024  & 2018/12/14          & 58466.2\\
        ${T_3}$           & FSRQ$\longrightarrow$BL Lac &58638-58664 & 2019/6/4-2019/6/30     &2.035$\pm$0.047  & 2019/6/25           & 58659.9\\
        ${B_2}$           & BL Lac                      &58665-58677 & 2019/7/1-2019/7/13     &1.897$\pm$0.052  & 2019/7/12           & $\ast$\\
       ${F_1}$-${B_2}$    & CL epoch                   &54628-58677 & 2008/6/11-2019/7/13    &2.234$\pm$0.022  & 2016/8/31-2019/7/12 &58171.9-58659.9\\
		\hline
	\end{tabular}
    \\{\footnotesize{
    Note. Column (1) gives the nomenclature of the epoch in CL; Column (2) indicates the states of CL blazar OQ 334 in various epochs; Columns (3)-(4) give the epoch records of $\gamma$-ray in CL observed by \textit{Fermi}-LAT, expressed in units MJD, year/month/day, respectively; Column (5) indicates the photon spectral index of $\gamma$-ray and their 1-$\sigma$ confidence interval; Columns (6) - (7) give the epoch records of the X-ray and optical/UV bands observed by Swift, respectively, which $\ast$ represent that there is data at this stage but not data in Swift, and $\star$ means that no corresponding data is found.}}
\end{table*}

\subsection{ Multi-wavelength SED }

\begin{figure*} 
    \centering
    \subfloat[\label{fig:a}][The FSRQ state in MJD 54628-58082]{
        \includegraphics[scale=0.43]{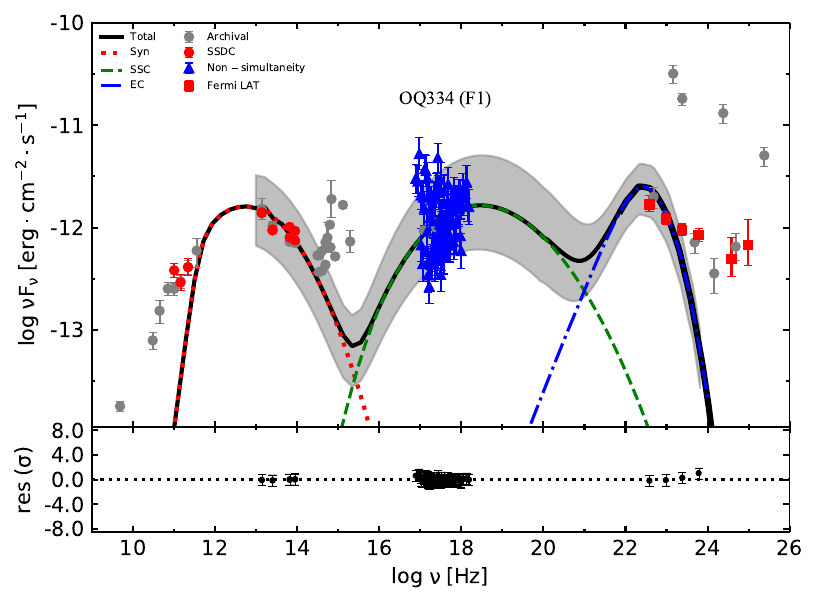}}\hspace{2mm}
    \subfloat[\label{fig:b}][FSRQ conversion to BL Lac state in MJD 58083-58121]{
        \includegraphics[scale=0.43]{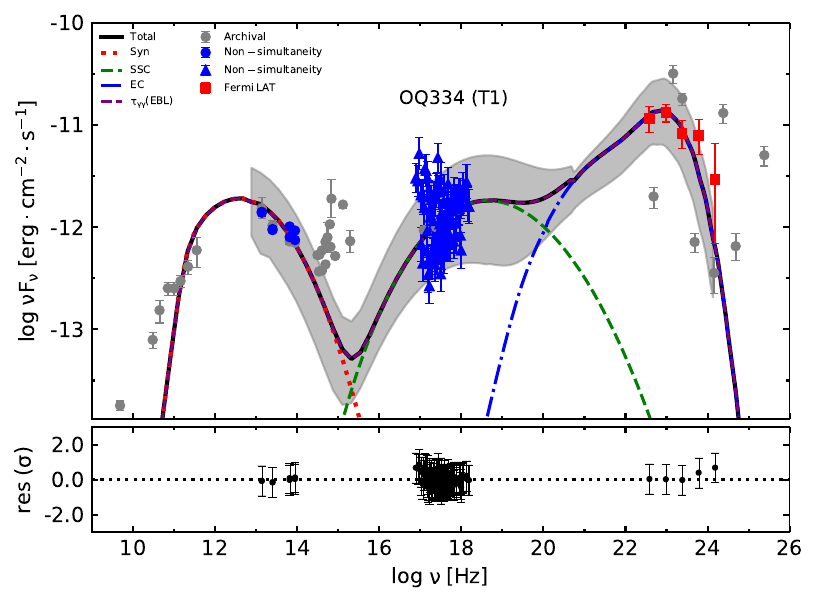}}
  
    \subfloat[\label{fig:c}][The BL Lac state in MJD 58122-58124]{
        \includegraphics[scale=0.43]{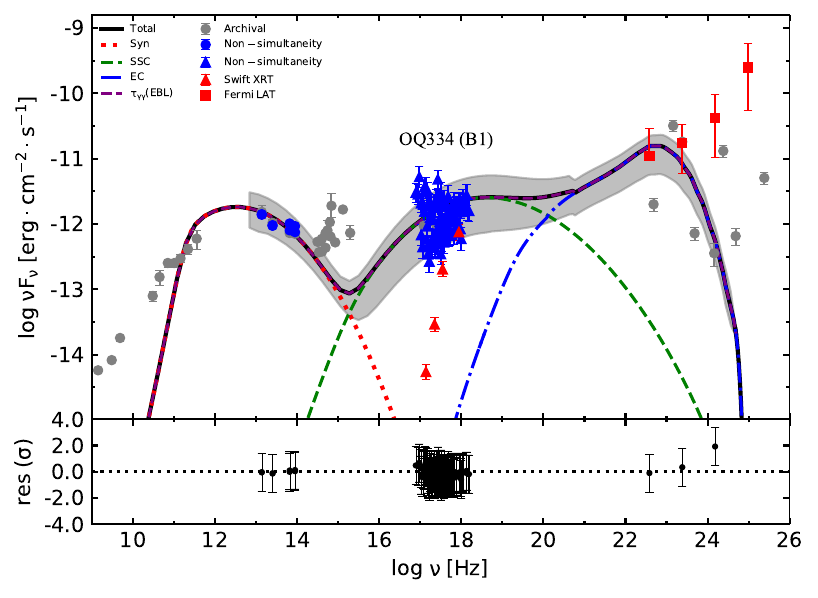}}\hspace{2mm}
    \subfloat[\label{fig:d}][BL Lac conversion to FSRQ state in MJD 58125-58220]{
        \includegraphics[scale=0.43]{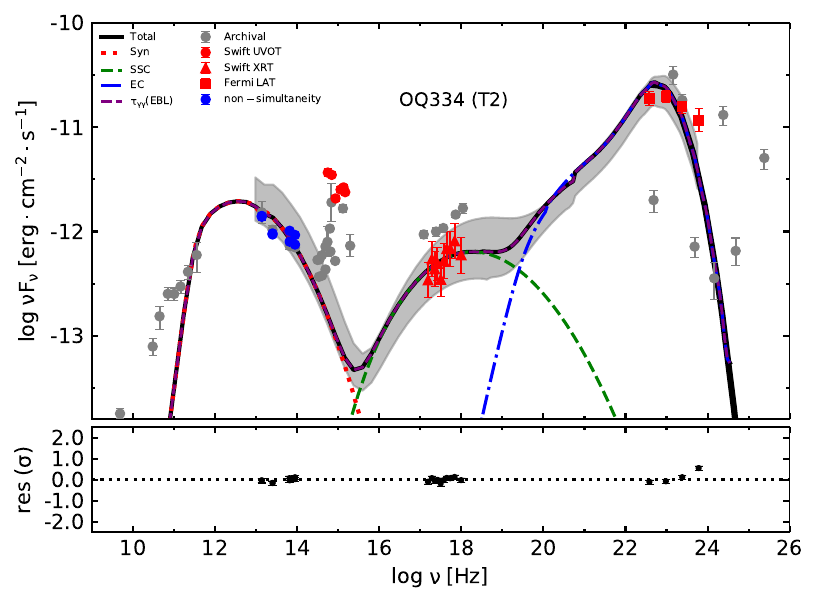} }
  
    \subfloat[\label{fig:subfig_e}][The FSRQ state in MJD 58221-58637]{
        \includegraphics[scale=0.43]{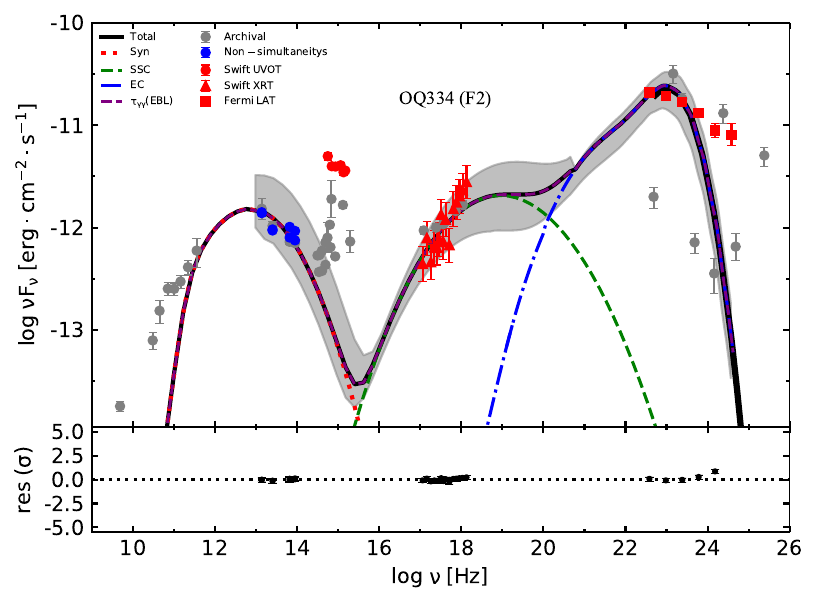}}\hspace{2mm}
    \subfloat[\label{fig:f}][FSRQ conversion to BL Lac state in MJD 58638-58664]{
        \includegraphics[scale=0.43]{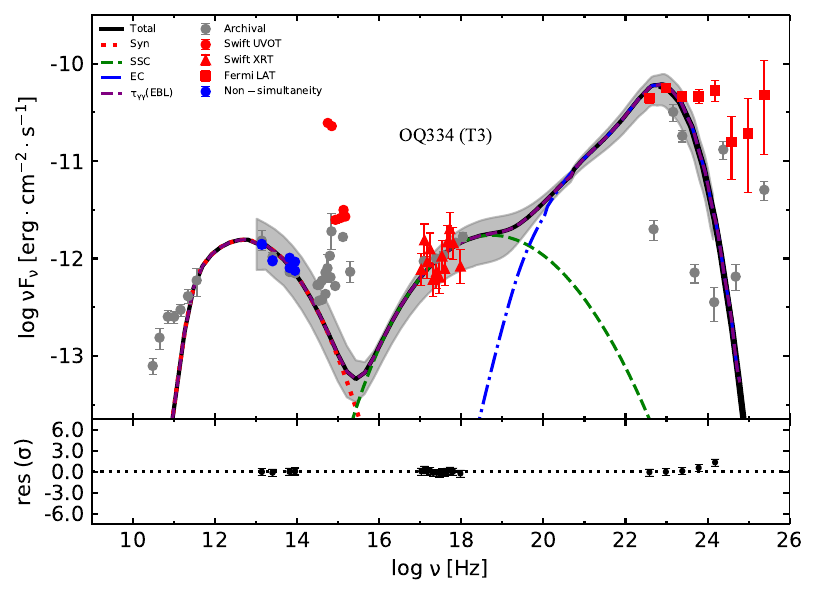}}

  \subfloat[\label{fig:g}][The BL Lac state in MJD 58665-58677]{
        \includegraphics[scale=0.43]{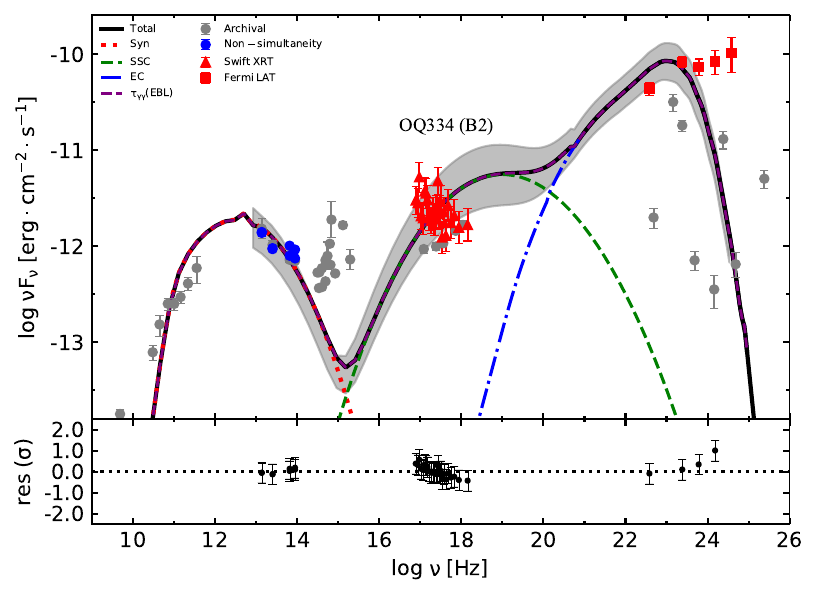}}\hspace{2mm}
    \subfloat[\label{fig:h}][The CL blazar state in MJD 54628-58677]{
        \includegraphics[scale=0.43]{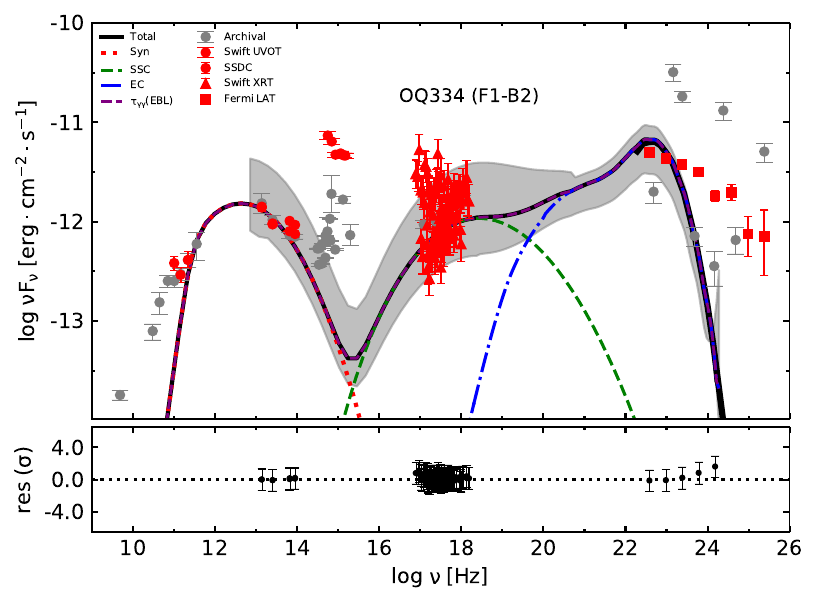}}

    \caption{Multi-wavelength SED of the CL phenomenon. The dotted red, dashed green, dot-dashed blue, and dashed purple lines represent the Syn, SSC, EC, and EBL absorption, respectively. The solid black line represents the total spectrum of the SED fit. The blue dots represent quasi-simultaneous data. Blue dots represent non-quasi-simultaneous data. Grey dots represent archival data. In these subgraphs, F1, T1, B1, T2, F2, T3, and B2 represent the SED in each state of the CL phenomenon, and F1-B2 represents the SED of the entire CL phenomenon. The grey background represents the 1$\sigma$ parameter space of the SED model and is displayed starting from the optical observation data points participating in the fit. The lower panel of each subplot shows the residual of the fit.}
    \label{Fig04} 
\end{figure*}

We fitted the multi-wavelength SED plot of OQ 334 during the CL phenomenon using a one-zone leptonic model (Syn+ SSC+ EC), as illustrated in Fig.~\ref{Fig04}.
To obtain quasi-simultaneous data for eight epochs and to better constrain the SED, we searched for the corresponding times and present them in Table~\ref{tab:01}.
For the ${F_1}$ epoch, we obtanied optical data points from the SSDC website; these are listed in Table~\ref{tab:04}.
However, no UV, optical, or X-ray data were available for the ${T_1}$ epoch, and therefore we included archival and non-quasi-simultaneous data to constrain the SED fitting plot for that particular epoch.

In order to constrain the peak value of synchrotron, we selected the optical data from the ${F_1}$ epoch as non-simultaneous data for epochs without quasi-simultaneous optical data (${T_1}$, ${B_1}$, ${T_2}$, ${F_2}$, ${T_3}$, and ${B_2}$ ). Additionally, it is generally considered that the radio emission region is relatively large compared to other bands, and therefore the radio data ($log\nu$ $<$ 11.5~Hz ) are not included in the fitting process \citep{2014ApJS..215....5K}.
In order to restrict SSC, for the ${F_1}$ and ${T_1}$ epochs, we use the X-ray data in their averaged state for these epochs. Specifically, this refers to the X-ray data obtained by averaging observations from the ${T_2}$, ${F_2}$, ${T_3}$, and ${B_2}$ epochs \citep{2021ApJ...915...59Z}.
In particular, for epochs ${B_1}$ and ${F_1}$-${B_2}$ , we continue to employ the averaged X-ray data to restrict SSC, as their quasi-simultaneous X-ray data exhibit significant discrepancies. Due to constraints imposed by the synchrotron process on the maximum electron energy, the emission flux in the very high-energy band does not meet expectations. As a result, data points with energies exceeding 10 GeV are excluded from all epochs of SED fitting.
Furthermore, it was observed that the archival data of OQ 334 exhibit a prominent `big blue bump' (14~Hz $<$ $log\nu$ $<$ 16~Hz). However, as this hump originates from thermal radiation of the DT, the accretion disc, and the host galaxy \citep{2001A&A...375..739D,2010ApJ...716...30A,2012MNRAS.420.2899G,2015ApJ...810...14A}, it was not possible to use data points within the region of the large blue bump to constrain the SED during the fitting process.
In Fig.~\ref{Fig04}, we employ the least-squares fitting technique \citep{2011ApJ...733...14M,2017ApJ...837...38K} to determine the 1$\sigma$ parameter space (confidence interval) in the SED model and represent it as a shaded gray region.
It is worth noting that the X-ray data for epochs ${F_1}$, ${T_1}$, ${B_1}$, and ${F_1}$-${B_2}$ are relatively discrete, and the fitting line for the upper and lower boundaries of the parameter space cannot reach 1$\sigma$.
Therefore, the X-ray data are not taken into account when giving the 1$\sigma$ parameter space of these four epochs, and this space should contain all the observed optical and $\gamma$-ray fitting data points whenever possible.
The determination of the 1$\sigma$ parameter space in epochs ${T_2}$, ${F_2}$, ${T_3}$, and ${B_2}$ involves taking into account optical, X-ray, and $\gamma$-ray observational data points.
Within this parameter space, the fitting line with the smallest chi-square value is selected as the locally optimal line for SED fitting. Additionally, the 1$\sigma$ uncertainty for each free parameter is determined using the standard deviation.
At the same time, the 1$\sigma$ uncertainty for each free parameter is given by the standard deviation (\citep{2012msma.book.....F,2020OJAp....3E...2P}; for example $\sigma = \sqrt{\frac{\sum_{i=1}^{n} (x_{i}-\bar{x})}{n}}$).

Local optimal parameter values obtained from the multi-wavelength SED fitting and the corresponding chi-square values $\chi ^{2}$ were recorded and are presented in Table~\ref{tab:05}.

\begin{figure}  
  \centering
   \includegraphics[width=90mm]{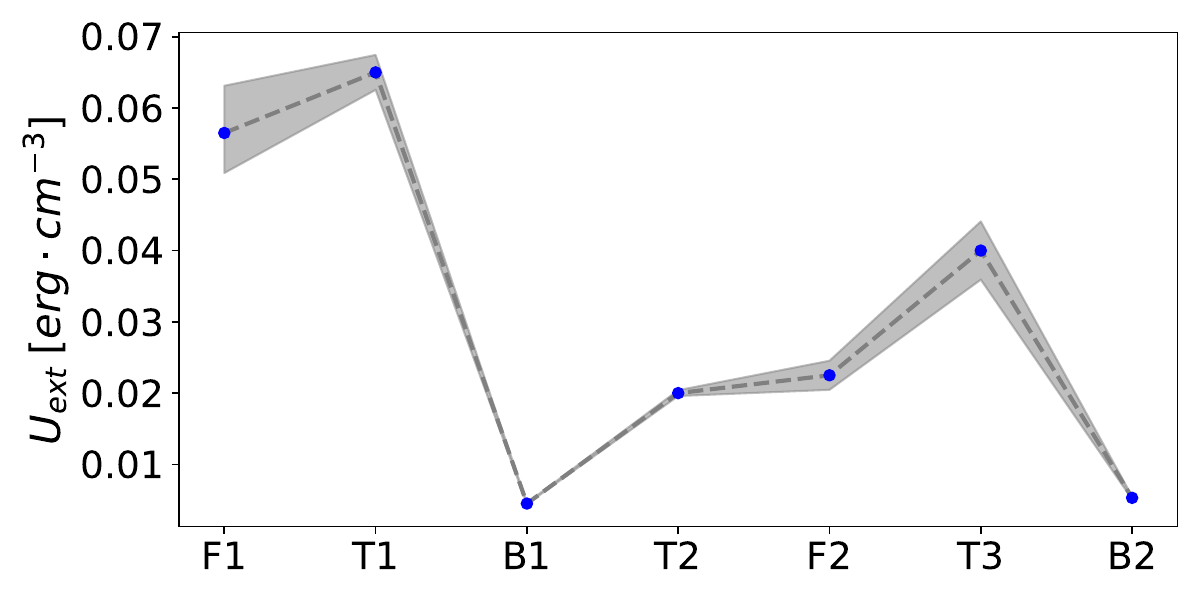}    
  \centering
   \caption{Variation in the energy density of the external photon field $U_{ext}$ in seven epochs in the CL. The grey background is a 1$\sigma$ confidence band of $U_{ext}$.}
   \label{Fig05}
\end{figure}

This fitting result reveals that the energy density of the external photon field varies across each epoch, with $U_{ext}$ being higher in epochs ${F_1}$ and ${F_2}$ compared to epochs ${B_1}$ and ${B_2}$.
This difference exhibits an evident oscillatory mode (as shown in Fig.~\ref{Fig05}).

\begin{table}
	\centering
	\caption{Time record of optical quasi-simultaneous data at the ${F_1}$ epoch.}
	\label{tab:04}
	\begin{tabular}{cccc} 
		\hline \hline
        Name  &  Start Time & Stop Time \\
		\hline
		${PCCS1F100}$ & 2009/8/12  & 2010/11/27 \\
		${PCCS1F143}$ & 2009/8/12 & 2010/11/27 \\
		${PCCS1F217}$ & 2009/8/12 & 2010/11/27 \\
        ${WISE ~W1 ~PointPsf}$ & 2010/1/12 & 2010/7/7 \\
        ${WISE ~W2 ~PointPsf}$ & 2010/1/12 & 2010/7/7 \\
        ${WISE ~W3 ~PointPsf}$ & 2010/1/12 & 2010/7/7 \\	
        ${WISE ~W4 ~PointPsf}$ & 2010/1/12 & 2010/7/7 \\
		${allwise~w1}$ & 2010/1/12 & 2011/1/10 \\
		${allwise~w2}$ & 2010/1/12 & 2011/1/10 \\
        ${allwise~w3}$ & 2010/1/12 & 2011/1/10 \\
        ${allwise~w4}$ & 2010/1/12 & 2011/1/10 \\
		\hline
	\end{tabular}
\end{table}


\section{Discussion}\label{sec:diss}
\subsection{\textit{Fermi}-LAT light curve and SED}
\subsubsection{Evolution of OQ 334}

By examining the light curve in Fig.~\ref{Fig02} and referring to Table~\ref{tab:01}, we observed that OQ 334 spends a duration of $t_{F_{1}}= 3453 $ days in the FSRQ state during the ${F_1}$ epoch, and $t_{F_{2}}= 416 $ days in the FSRQ state during the ${F_2}$ epoch.
During ${F_1}$, OQ 334 remains in quiescence, while it becomes active during the ${F_2}$ epoch.
On the other hand, OQ 334 enters the BL Lac state for $t_{B_{1}}= 2 $ days during the ${B_1}$ epoch, and $t_{B_{2}}=12 $ days during the ${B_2}$ epoch, with frequent flux variations observed.
OQ 334 is in the FSRQ state for a shorter epoch, and is in the BL Lac state for longer.
As evident from the oscillations in the EW changes shown in Fig.~\ref{Fig01}, OQ 334 transitions back to a FSRQ state after the ${B_2}$ epoch, and the duration of its FSRQ state is expected to be shorter than the ${F_2}$ epoch (416 days).
With continuous oscillatory evolution, OQ 334 may eventually evolve into a BL Lac object \citep{2023arXiv231005096P}.
Changing-look AGNs represent an intermediate class of transitional galaxies between low-luminosity active galaxies and high-luminosity quasars \citep{2022ApJ...927..227L}, and CL blazars may be a class of excessive blazar sources in transition from FSRQ to BL Lac type \citep{2024ApJ...962..122K}.
Follow-up spectroscopic observations are anticipated; these should provide further evidence to support this hypothesis.
\subsubsection{$\gamma$-ray flux fluctuation}
The values of fractional flux variability indicate that the flux variation amplitude in the high-energy band is larger than it is in the low-energy band. On the other hand, phenomenologically, it can be seen from Fig.~\ref{Fig02} that the flux variation frequency in the high-energy band less than it is in the low-energy band. These scenarios suggest that the level of variability is very high during the CL epoch.

When high-energy $\gamma$-rays propagate through the Universe, they experience energy loss due to absorption by cosmic background light \citep{2001ApJ...553..538W}.
The degree of absorption increases with the redshift of the blazar.
However, in the case of the CL phenomenon seen in OQ 334, the redshift value is 0.6819 \citep{2010MNRAS.405.2302H}, which is a constant, allowing us to disregard the influence of cosmic background light in the explanation of why the low-energy band exhibits more pronounced fluctuations than the high-energy band.
These latter authors also identified that seed photons in the CL phenomenon mainly originate from the BLR, interacting with relativistic electrons in the radiation region to produce the primary $\gamma$-ray.
However, the diffuse soft photon field generated by the BLR absorbs the high-energy $\gamma$-ray (GeV) \citep{2006ApJ...653.1089L,2008ApJ...688..148L}, resulting in fewer $\gamma$-rays being observed in the high-energy band and a relatively reduced flux variation.

\subsubsection{Changes in photon spectral index $\Gamma_{\gamma}$}
Values of the photon spectral index $\Gamma_{\gamma}$ in Table~\ref{tab:01}  were obtained through \textit{Fermipy} analysis.
The $\gamma$-ray photon spectral index during the ${F_1}$ and ${F_2}$ epochs has been found to exceed 2, while it is below 2 during the ${B_1}$ and ${B_2}$ epochs.
This is consistent with the results of \cite{2010ApJ...710.1271A} regarding the $\gamma$-ray spectral index results of FSRQ and BL Lac.
The separation of hard (BL Lac) and soft (FSRQ) spectra in OQ 334 during EW changes from ${F_1}$ to ${B_2}$ can be attributed to the significantly different cooling radiation of relativistic electrons in the jet \citep{1998MNRAS.297..348G,2009MNRAS.396L.105G}.
For FSRQ objects, $\gamma$-ray radiation is primarily dominated by the EC radiation mechanism \citep{2021Ap&SS.366...12O}.
It is possible that there exists a similar EC component in the BL Lac state leading ${F_1}$-${B_2}$ to exhibit the photon spectral index characteristic of FSRQ.
Therefore, the variation in soft photons involved in EC radiation might be the primary cause of the observed spectral index hardening or softening.

\subsection{Multi-wavelength SED}
\subsubsection{External photon field energy density and accretion rate}
Generally, when the EW is less than 5 {\AA} in BL Lac objects, these latter exhibit neither strong emission lines nor a significant EC component. Nevertheless, there are a few BL Lacs that necessitate the inclusion of weak EC components to resolve the extreme parameter problems associated with SED fitting \citep{2008ApJ...684L..73A,2011ApJ...726...43A,2014ApJ...783...83L,2015RAA....15..313L,2016MNRAS.461.1862K}.
In some BL Lacs, the presence of weak emission lines has been detected, suggesting the existence of EC components \citep{2011ApJ...732..113S,2016ApJ...832...76D}.
Furthermore, \cite{2021ApJ...913..146M} showed the presence of weak emission lines in the BL Lac state, implying that the EC component of OQ 334 may be present in the BL Lac state (${B_1}$, ${B_2}$), albeit it weaker than in the FSRQ state.

The CL phenomenon of OQ 334 was studied using the one-zone leptonic model Syn+ SSC+ EC.
The energy density of the external photon field was fitted as presented in Table~\ref{tab:05}, along with the corresponding fitting plot shown in Fig.~\ref{Fig04}.
Our SED fitting revealed $K_{0} \approx 1\times 10^{-5}$ in the eight epochs, and so the $K_{0}$ is no longer used as a free parameter, and therefore dof = 7 in this paper.
As the solid black line in the SED plot is the fitted line with the smallest chi-square value in the 1$\sigma$ confidence band, the magnitude of the chi-square value is related to the distance from the observed data point to the fitted line.
Moreover, as can be seen from the lower panel of each subplot in Fig.~\ref{Fig04}, the chi-square value is smaller when there are more data points near the black line (that is the local best-fit line).
This results in the $\nu$$f_{\nu}$ value of the fitted line near the peak of the EC component of (d), (e), and (h) in Fig.~\ref{Fig04} being higher than the $\nu$$f_{\nu}$ value of the $\gamma$-ray data point at the same frequency.

\cite{1998MNRAS.301..451G} compared the SSC and EC models, considering that seed photons in EC radiation are provided by Syn radiation both within and external to the jet.
These authors concluded that the EC scattering process becomes unimportant when the energy density of EC scattering is lower than the energy density of Syn radiation.
In this scenario, SSC radiation would be stronger than EC radiation, and the effect of EC radiation might even be negligible.
This corresponds to the SED feature of the BL Lac state. However, in our fitted results (Fig.~\ref{Fig04}), the SSC of $F_{1}-B_{2}$ does not appear to be stronger than the EC. This may be attributed to the fact that OQ 334 also has a weaker EC component when in the BL Lac state, which results in the SED exhibiting FSRQ characteristics throughout the CL event.

The origin of the external photon field is closely related to the location of the $\gamma$-ray radiation region.
The source of seed photons in the external photon field can be limited by the size of the Schwarzschild radius, $R_{S}$, and the location of the radiation region, $r_{dis}$, where ${R_{S}=2GM/c^{2}}$.
When $r_{dis}$ is smaller than a few hundred $R_{S}$, seed photons are derived from the accretion disc.
If $r_{dis}$ is larger than a few hundred $R_{S}$ but smaller than a few thousand $R_{S}$, the seed photons originate from the BLR.
When $r_{dis}$ exceeds the radius of the outer boundary of the BLR and is less than $\sim 10^{5}R_{S}$, the seed photons come from the DT.
Finally, when $R > 10^{5}R_{S}$, the seed photons are derived from cosmic microwave background radiation \citep{2009MNRAS.397..985G,2014ApJS..215....5K}.
The accretion disc luminosity, $L_{disc} = 9.200 \times 10^{45} erg s^{-1}$ , and the black hole mass, $M_{SMBH}=3.980\times 10^{8} M_{\odot}$ (where $M_{\odot}$ is the solar mass with a magnitude of $1.989\times 10^{33}$ g), were calculated based on the work of \cite{2021MNRAS.502.5245P}.
Additionally, the Schwarzschild radius is determined as $R_{S} \approx 1.175\times 10^{14} cm $.
In this case, the value of several hundred $R_{S}$ is approximately $\sim 10^{2}R_{S} \simeq 10^{2}\times R_{S} \approx 1.175\times 10^{16} cm $, and the value of several thousand $R_{S}$ is approximately $\sim 10^{3}R_{S} \simeq 10^{3} \times R_{S} \approx 1.175 \times 10^{17} cm $.
Based on our scenario, the fact that $r_{dis}$ satisfies $10^{2}R_{S} < r_{dis} < 10^{3}R_{S}$ (as indicated in Table~\ref{tab:05}) implies that the seed photons of the external photon field during every epoch are primarily provided by the BLR.
The energy density of the external photon field is approximately equal to the energy density of the BLR, $U_{ext}\sim U_{BLR}$.


\begin{table*}
	\centering
    \renewcommand\arraystretch{2}
    \tabcolsep=0.12cm
	\caption{Multi-wavelength SED fits to local optimal parameters and the chi-square value $\chi ^{2}$.}
	\label{tab:05}
	\begin{tabular}{ccccccccccc} 
		\hline \hline
         Epoch &\makebox[0.01\textwidth][c]{s}  &\makebox[0.01\textwidth][c]{b}  &\makebox[0.01\textwidth][c]{$\gamma_{0}$}  &\makebox[0.01\textwidth][c]{$\delta$} &\makebox[0.01\textwidth][c]{B} & \makebox[0.01\textwidth][c]{$R$}& \makebox[0.01\textwidth][c]{$r_{dis}$} &\makebox[0.01\textwidth][c]{$U_{ext}$}& \makebox[0.01\textwidth][c]{$\chi^{2}$} \\
         ...   &...&...& $10^{3}$ &...&...& $10^{16}$ & $10^{16}$ & $10^{-2}$ & ...\\
         ...   &...&...          & ...&...& G & cm& cm & $erg\cdot cm^{-3}$ &... \\
		\hline
        ${F_1}$        & 5.400$\pm$0.122 &1.175$\pm$0.038  &7.200$\pm$0.120    &7.200$\pm$0.122  &1.650$\pm$0.367 &1.109$\pm$0.019 &1.241$\pm$0.123  & 5.700$\pm$0.600 &8.432  \\
		${T_1}$        & 5.240$\pm$0.033 &1.175$\pm$0.017  &7.450$\pm$0.321    &10.500$\pm$0.448 &0.800$\pm$0.082 &1.617$\pm$0.069 &1.180$\pm$0.049 & 6.500$\pm$0.200 &7.282   \\
		${B_1}$        & 5.065$\pm$0.135 &1.020$\pm$0.027  &7.400$\pm$0.060    &12.400$\pm$0.408 &0.270$\pm$0.008 &1.910$\pm$0.063 &2.500$\pm$0.082 & 0.450$\pm$0.010 &8.162  \\
		${T_2}$        & 5.300$\pm$0.082 &1.110$\pm$0.048  &4.900$\pm$0.121    &15.340$\pm$0.220 &0.600$\pm$0.020 &2.362$\pm$0.034 &1.530$\pm$0.074 & 2.000$\pm$0.040 &5.271   \\
        ${F_2}$        & 5.770$\pm$0.029 &1.445$\pm$0.038  &7.550$\pm$0.281    &13.250$\pm$0.448 &0.380$\pm$0.045 &2.041$\pm$0.069 &1.370$\pm$0.045 & 2.250$\pm$0.200 &6.211  \\
        ${T_3}$        & 5.295$\pm$0.012 &1.100$\pm$0.021  &7.550$\pm$0.040    &13.950$\pm$0.285 &0.390$\pm$0.033 &2.148$\pm$0.044 &1.400$\pm$0.082 & 4.000$\pm$0.410 &18.663   \\
        ${B_2}$        & 5.110$\pm$0.118 &1.255$\pm$0.040  &8.600$\pm$0.201    &12.200$\pm$0.135 &0.092$\pm$0.002 &1.879$\pm$0.021 &5.200$\pm$0.102 & 0.530$\pm$0.020 &7.932  \\
        ${F_1}$-${B_2}$ & 5.500$\pm$0.082 &1.230$\pm$0.010  &5.600$\pm$0.060   &9.200$\pm$0.167  &0.800$\pm$0.163 &1.417$\pm$0.026 &1.900$\pm$0.041 & 2.400$\pm$0.200 &6.853  \\
      \hline
      \end{tabular}
      \\
      {\footnotesize{
       Note. $R=ct_{var}\delta/(1+z)$, $t_{var}$ is the minimum light curve time scale. The columns in the table represent the following:1. Epoch, the state of OQ 334 during different epochs within the CL phenomenon; 2. $s$, the spectral index; 3. $b$, the spectral curvature; 4. $\gamma_{0}$, the initial Lorentz factor; 5. $\delta$, the Doppler factor; 6. B, the magnetic field, in units of Gs; 7. $R$, the radius of the radiation zone, specified in $cm$; 8. $r_{dis}$, the location of the radiation zone, specified in $cm$; 9. $U_{ext}$, the energy density of the external photon field, denoted in units of $erg\cdot cm^{-3}$; 10. $\chi ^{2}$, the chi-square value.}}	
\end{table*}

The energy density of the BLR under the observer coordinate system can be approximated as \citep{2008MNRAS.387.1669G}
\begin{large}
\begin{equation}
    \begin{array}{l}
    U_{BLR} \approx \frac{0.1L_{disc}}{4\pi R_{BLR}^{2}c}\;\;\;\;\;\;\;\;\;\;\;\;\;\;\;\;  r_{dis} < R_{BLR}
    \end{array}
    \label{Eq:6}
\end{equation}
\end{large}

where $L_{disc}$ represents the accretion disc luminosity and $R_{BLR}$ is the characteristic radius of the BLR, and is given by $R_{BLR}=10^{17} L_{disc,45}^{1/2}\approx 3.033\times 10^{17} cm$.
According to \cite{2017ApJS..228....1Z}, the radius of the outer boundary of the BLR is $R_{BLR,out}=2.7R_{BLR}\approx 8.190\times 10^{17} cm$.
We observed that the location of the radiation zone for all epochs obtained through fitting satisfies $r_{dis}$ < $R_{BLR,out}$.
Consequently, {$U_{ext}{\sim}U_{BLR}\approx{\frac{0.1L_{disc}}{4\pi R_{BLR}^{2}c}}$ holds more credibility.
By using} the energy density of the external photon field, we can estimate the accretion disc luminosity (see Table~\ref{tab:06}).
In the co-moving coordinate system, the energy density of the external photon field is:
\begin{large}
\begin{equation}
    \begin{array}{l}
    U_{ext}^{\prime}\approx\frac{17\Gamma^{2}}{12}U_{ext}\approx
    \frac{17\Gamma^{2}}{12}\cdot\frac{0.1L_{disc}}{4\pi R_{BLR}^{2}c}
    \end{array}
    \label{Eq:7}
\end{equation}
\end{large}
where $L_{disc}$ is the Lorentz invariant, which is consistent in the co-moving coordinate system and the observer coordinate system.
We used the accretion disc luminosity to estimate the accretion rate \.{m}= \.{M}/\.{M}$_{Edd}$ \citep{2003ApJ...599..147C}, where the Eddington luminosity is given by $L_{Edd}=1.257\times 10^{38}\frac{M}{M_{\odot}} erg~s^{-1}$ \citep{1980ApJ...242..772A}, and \.{M}$_{Edd}$=$L_{Edd}/\eta c^{2}$.
The accretion disc luminosity can be expressed as $L_{disc}$ = $\eta$\.{M}$c^{2}$ \citep{2006smqw.confE..27G}, leading to \.{m}=$L_{disc}/L_{Edd}$, as shown in Table~\ref{tab:06}.

\subsubsection{Accretion mode and the CL phenomenon}
Active galactic nuclei are intimately linked to the accretion of matter onto SMBHs.
The accretion process occurs on the accretion disc, generating a continuous stream of photons that ionise and excite the gas in the BLR, leading to the formation of broad emission lines \citep{1997iagn.book.....P}.
The instability of the accretion disc produces changes in the radiation emitted from the disc \citep{1986ApJ...305...28L}, resulting in changes in the continuous photons and subsequently impacting the generation of broad emission lines.
Changes in the accretion rate can alter the structure of the BLR, and the BLR may even vanish at low accretion rates \citep{2006ApJ...648L.101E}.
The accretion disc model provides a good explanation for the brightness and compactness of AGNs.
Generally, high-luminosity (high-accretion-rate) AGNs exhibit geometrically thin, optically thick standard accretion discs (i.e. Shakura-Sunyaev discs; SSDs, \citealt{1973A&A....24..337S}).
However, in AGNs with low luminosity (LLAGNs) and low accretion rates \citep{2014ARA&A..52..529Y}), the SSD may transition to an advection-dominated accretion flow (ADAF, \citep{1995ApJ...444..231N,1996ApJ...462..142L,1999ApJ...516..177G}) which can account for most of the observed features of LLAGNs.
In LLAGNs, when the gas-outflow temperature of the ADAF is excessively high, it cannot be efficiently cooled within the BLR, leading to inhibiting cloud formation and ultimately resulting in the disappearance of the BLR \citep{2010ApJ...724..855C}.

The variation in the oscillation of the EW implies that there must have been a significant change between the absorption or emission line profile and the continuous spectrum during the CL event.
This strongly suggests that CL events caused by obscurations can be largely ruled out.
In this scenario, a large number of photons from the broad emission lines are fed back into the IC process and scattered into the $\gamma$-ray band \citep{2022ApJ...936..146X}.
Therefore, the strength of the emission lines directly influences the number of seed photons available for the IC process.
Consequently, the EW of a BL Lac object, influenced by an ADAF \citep{2002ApJ...579..554W,2003ApJ...599..147C, 2014ARA&A..52..529Y,2021Ap&SS.366...12O}, is expected to be smaller than that of a FSRQ object, which is influenced by an SSD.

Our analysis revealed that the accretion rate of OQ 334 was \.{m}>0.1 during its FSRQ (${F_1}$, ${F_2}$) states, whereas it was \.{m} $<$ 0.1 during its BL Lac (${B_1}$, ${B_2}$) states. This is similar to that BL Lacs and FSRQs can be distinguished based on luminosity ($L_{BLR}/L_{Edd} =10^{-3}$, that is \.{M}/\.{M}$_{Edd}$ =0.1, \citealt{2010MNRAS.402..497G,2011MNRAS.414.2674G,2012MNRAS.421.1764S,2014MNRAS.441.3375X}).  
Our results are similar to the findings of \cite{2019MNRAS.482L..80B} on accretion rate differentiation between BL Lacs and FSRQs.
Concerning the mass-accretion rate and blazar classification: \.{m} > 0.1 for FSRQs, which corresponds to a disc with high radiation efficiency; and for BL Lac objects, \.{m} $<$ 0.1, indicating an ADAF.

The results of the calculations (see Table~\ref{tab:06}) that the accretion rate exhibits oscillatory variations, and we assume that the transition from FSRQ to BL Lac represents a state change, which may be influenced by the variation of accretion rate. This supports the hypothesis that the accretion disc is in a state of active oscillatory change.
During the transition from FSRQ to BL Lac type, the disc may be in the state of transitioning SSD to ADAF, but it is not completely converted into an ADAF disc.
That is, this state should be in a narrow region where the internal ADAF and the external SSD disc are connected \citep{2020A&A...641A.167S,2023NatAs...7.1282R}.
Usually, FSRQ corresponds to an SSD disc and BL Lac corresponds to a disc in the ADAF state \citep{2002ApJ...571..226C,2002ApJ...570L..13C,2002ApJ...579..554W,2003ApJ...599..147C,2014ARA&A..52..529Y,2019MNRAS.482L..80B,2021Ap&SS.366...12O}.
If the SSD disc is to be fully converted into an ADAF disc, a long-term change is necessary for completion.
However, due to the source-current-unstable transitional state, the transition of the state of the disc and the change of accretion rate may occur on a short timescale.
During the transition from FSRQ to BL Lac state, the EC component may weaken but should not disappear completely, in other words, BLR clouds can only weaken and cannot be destroyed.

When the accretion rate \.{m} exceeds 0.1 (${F_1}$, ${F_2}$) and a geometrically thin, optically thick SSD is present, the accretion material effectively reprocesses and scatters the radiation from the accretion disc.
As a result, the strong optical emission line is observed in the BLR \citep{2002ApJ...564...86B}.
In this SSD accretion mode, the optical/UV emission line is enhanced, leading to a larger EW and the seed photons increase during the IC process. Consequently, the energy density of the external photon field increases, indicating that OQ 334 is in the FSRQ state.

On the other hand, when the accretion rate \.{m} drops below 0.1 (${B_1}$, ${B_2}$), and the ADAF state is functioning, the photons capable of ionising a broad-emission-line gas cloud decrease in number \citep{2022Galax..10...35P}.
This results in the suppression of the optical/UV emission line in the ADAF accretion mode.
The ionisation of the broad-spectral-line gas cloud becomes limited, or only a small portion is ionised.
Consequently, the seed photons available for the IC process are relatively reduced, leading to a smaller EW.
This reduction in seed photons causes a decrease in the energy density of the external photon field.
Therefore, during this regime, OQ 334 is in the BL Lac state.
It is worth noting that the radiation region $r_{dis}$ also plays a role in affecting the energy density of the external photon field, $U_{ext}$, when the accretion rate changes \citep{1996MNRAS.280...67G,2009MNRAS.397..985G,2017ApJ...837...38K}, though the specific physics governing the inverse correlation between $r_{dis}$ and $U_{ext}$ has not been addressed in this work.

Therefore, we conclude that the accretion disc in OQ 334 functions in SSD accretion mode during the FSRQ state, but operates in ADAF accretion mode during the BL Lac state. The detailed physics governing the transition between these accretion modes remains unclear. However, this transition leads to changes in the accretion rate, resulting in variations in the accretion disc luminosity. Consequently, the energy density of the external photon field is altered, corresponding to changes in the EW of OQ 334. The observed CL phenomenon may be attributed to these changes in the accretion mode \citep{2024ApJ...962..122K}, suggesting a strong connection between the accretion mode and the variability of the source.

\begin{table}
	\centering
	\caption{Accretion disc luminosity $L_{disc}$ and accretion rate \.{m} over eight epochs in the CL phenomenon.}
	\label{tab:06}
	\begin{tabular}{cccc} 
		\hline \hline
        \makebox[0.01\textwidth][c]{Epoch}  & State                        & $L_{disc}$           & \.{m}   \\
        ...    & ...                          & $10^{45}$           &  ...    \\
        ...    & ...                          & $erg\cdot s^{-1}$           &  ...    \\
		\hline
        ${F_1}$ & FSRQ                        & 19.756$\pm$2.080 & 0.395$\pm$0.042 \\
		${T_1}$ & FSRQ $ \rightarrow $ BL Lac & 22.528$\pm$0.693 & 0.450$\pm$0.014  \\
		${B_1}$ & BL Lac                      & 1.560$\pm$0.035  & 0.031$\pm$0.001   \\
		${T_2}$ & BL Lac $ \rightarrow $ FSRQ & 6.932$\pm$0.139  & 0.139$\pm$0.003  \\
        ${F_2}$ & FSRQ                        & 7.798$\pm$0.693  & 0.156$\pm$0.014  \\
        ${T_3}$ & FSRQ $ \rightarrow $ BL Lac & 13.864$\pm$1.421 & 0.277$\pm$0.028  \\
        ${B_2}$ & BL Lac                      & 1.837$\pm$0.069  & 0.037$\pm$0.001 \\
        ${F_1}$-${B_2}$ & CL epoch            & 8.318$\pm$0.693  & 0.166$\pm$0.014 \\
		\hline
	\end{tabular}
    \\{\footnotesize{
    Note. The uncertainty of $L_{disc}$ and \.{m} is determined by the error transfer formula \citep{2012msma.book.....F}, $\sigma(y)=\sqrt{(\Sigma_{i=1}^{m}(\frac{\partial y}{\partial x_{i}})^2)(\sigma(x_{i}))^2 + 2\Sigma_{i,j=1,i\neq j}^{m}\frac{\partial y}{\partial x_{i}}\frac{\partial y}{\partial x_{j}}\rho_{ij}\sigma(x_{i})\sigma(x_{j})}$ , where $\rho_{ij}$ represent the correlation coefficient.}}
\end{table}
\section{Conclusion}    
The CL phenomenon of OQ 334 during the MJD 54628-58677 epoch can be separated into three distinct states, that is, the FSRQ state (${F_1}$, ${F_2}$), the BL Lac state (${B_1}$, ${B_2}$), and the transition state (${T_1}$, ${T_2}$, ${T_3}$), as shown in Table~\ref{tab:01}.
We analysed the \textit{Fermi}-LAT light curve and corresponding SED during these seven epochs, and studied the quasi-simultaneous multi-wavelength SED in each epoch using the Syn+SSC+EC model.

Our findings can be summarised as follows:
(1) During these eight epochs, the seed photons of the external photon field in OQ 334 are primarily sourced from the BLR.
(2) The change in the energy density of the external photon field is attributed to variations in the luminosity of the accretion disc, which are caused by transitions in the accretion mode.
(3) The accretion mode transition inhibits or enhances the continuous spectrum and optical/UV emission lines, resulting in the transition of OQ 334 between the FSRQ state and the BL Lac state.

In summary, the results strongly indicate that the CL phenomenon in OQ 334 is closely associated with variations in the accretion mode, which lead to significant changes in the observed flux, energy density, and spectral characteristics of the source across different epochs.



\section*{Acknowledgments}
We thank the anonymous editor and referee for very constructive and helpful comments and suggestions.
We acknowledge the use of data, analysis tools, and services from the Open Universe platform, the ASI Space Science Data Center (SSDC), the Astrophysics Science Archive Research Center (HEASARC), the Fermi Science Tools, the Astrophysics Data System (ADS), and the National Extra-galactic Database (NED).
This work is partially supported by the National Natural Science Foundation of China (Grant Nos. 12363002, 12163002, 12233007 and U1931203) and National SKA Program of China (2022SKA0120101).


\bibliographystyle{aa}
\bibliography{refer_oq334}

\begin{thebibliography}{144}
\expandafter\ifx\csname natexlab\endcsname\relax\def\natexlab#1{#1}\fi

\bibitem[{{Abdo} {et~al.}(2010{\natexlab{a}}){Abdo}, {Ackermann}, {Agudo}, {Ajello}, {Aller}, {Aller}, {Angelakis}, {Arkharov}, {Axelsson}, {Bach}, {Baldini}, {Ballet}, {Barbiellini}, {Bastieri}, {Baughman}, {Bechtol}, {Bellazzini}, {Benitez}, {Berdyugin}, {Berenji}, {Blandford}, {Bloom}, {Boettcher}, {Bonamente}, {Borgland}, {Bregeon}, {Brez}, {Brigida}, {Bruel}, {Burnett}, {Burrows}, {Buson}, {Caliandro}, {Calzoletti}, {Cameron}, {Capalbi}, {Caraveo}, {Carosati}, {Casandjian}, {Cavazzuti}, {Cecchi}, {{\c{C}}elik}, {Charles}, {Chaty}, {Chekhtman}, {Chen}, {Chiang}, {Chincarini}, {Ciprini}, {Claus}, {Cohen-Tanugi}, {Colafrancesco}, {Cominsky}, {Conrad}, {Costamante}, {Cutini}, {D'ammando}, {Deitrick}, {D'Elia}, {Dermer}, {de Angelis}, {de Palma}, {Digel}, {Donnarumma}, {Silva}, {Drell}, {Dubois}, {Dultzin}, {Dumora}, {Falcone}, {Farnier}, {Favuzzi}, {Fegan}, {Focke}, {Forn{\'e}}, {Fortin}, {Frailis}, {Fuhrmann}, {Fukazawa}, {Funk}, {Fusco}, {G{\'o}mez}, {Gargano}, {Gasparrini}, {Gehrels}, {Germani},
  {Giebels}, {Giglietto}, {Giommi}, {Giordano}, {Giuliani}, {Glanzman}, {Godfrey}, {Grenier}, {Gronwall}, {Grove}, {Guillemot}, {Guiriec}, {Gurwell}, {Hadasch}, {Hanabata}, {Harding}, {Hayashida}, {Hays}, {Healey}, {Heidt}, {Hiriart}, {Horan}, {Hoversten}, {Hughes}, {Itoh}, {Jackson}, {J{\'o}hannesson}, {Johnson}, {Johnson}, {Jorstad}, {Kadler}, {Kamae}, {Katagiri}, {Kataoka}, {Kawai}, {Kennea}, {Kerr}, {Kimeridze}, {Kn{\"o}dlseder}, {Kocian}, {Kopatskaya}, {Koptelova}, {Konstantinova}, {Kovalev}, {Kovalev}, {Kurtanidze}, {Kuss}, {Lande}, {Larionov}, {Latronico}, {Leto}, {Lindfors}, {Longo}, {Loparco}, {Lott}, {Lovellette}, {Lubrano}, {Madejski}, {Makeev}, {Marchegiani}, {Marscher}, {Marshall}, {Max-Moerbeck}, {Mazziotta}, {McConville}, {McEnery}, {Meurer}, {Michelson}, {Mitthumsiri}, {Mizuno}, {Moiseev}, {Monte}, {Monzani}, {Morselli}, {Moskalenko}, {Murgia}, {Nestoras}, {Nilsson}, {Nizhelsky}, {Nolan}, {Norris}, {Nuss}, {Ohsugi}, {Ojha}, {Omodei}, {Orlando}, {Ormes}, {Osborne}, {Ozaki}, {Pacciani},
  {Padovani}, {Pagani}, {Page}, {Paneque}, {Panetta}, {Parent}, {Pasanen}, {Pavlidou}, {Pelassa}, {Pepe}, {Perri}, {Pesce-Rollins}, {Piranomonte}, {Piron}, {Pittori}, {Porter}, {Puccetti}, {Rahoui}, {Rain{\`o}}, {Raiteri}, {Rando}, {Razzano}, {Reimer}, {Reimer}, {Reposeur}, {Richards}, {Ritz}, {Rochester}, {Rodriguez}, {Romani}, {Ros}, {Roth}, {Roustazadeh}, {Ryde}, {Sadrozinski}, {Sadun}, {Sanchez}, {Sander}, {Saz Parkinson}, {Scargle}, {Sellerholm}, {Sgr{\`o}}, {Shaw}, {Sigua}, {Siskind}, {Smith}, {Smith}, {Spandre}, {Spinelli}, {Starck}, {Stevenson}, {Stratta}, {Strickman}, {Suson}, {Tajima}, {Takahashi}, {Takahashi}, {Takalo}, {Tanaka}, {Thayer}, {Thayer}, {Thompson}, {Tibaldo}, {Torres}, {Tosti}, {Tramacere}, {Uchiyama}, {Usher}, {Vasileiou}, {Verrecchia}, {Vilchez}, {Villata}, {Vitale}, {Waite}, {Wang}, {Winer}, {Wood}, {Ylinen}, {Zensus}, {Zhekanis}, \& {Ziegler}}]{2010ApJ...716...30A}
{Abdo}, A.~A., {Ackermann}, M., {Agudo}, I., {et~al.} 2010{\natexlab{a}}, \apj, 716, 30

\bibitem[{{Abdo} {et~al.}(2010{\natexlab{b}}){Abdo}, {Ackermann}, {Ajello}, {Antolini}, {Baldini}, {Ballet}, {Barbiellini}, {Bastieri}, {Bechtol}, {Bellazzini}, {Berenji}, {Blandford}, {Bloom}, {Bonamente}, {Borgland}, {Bouvier}, {Bregeon}, {Brez}, {Brigida}, {Bruel}, {Buehler}, {Burnett}, {Buson}, {Caliandro}, {Cameron}, {Caraveo}, {Carrigan}, {Casandjian}, {Cavazzuti}, {Cecchi}, {{\c{C}}elik}, {Chekhtman}, {Cheung}, {Chiang}, {Ciprini}, {Claus}, {Cohen-Tanugi}, {Cominsky}, {Conrad}, {Costamante}, {Cutini}, {Dermer}, {de Angelis}, {de Palma}, {Silva}, {Drell}, {Dubois}, {Dumora}, {Farnier}, {Favuzzi}, {Fegan}, {Focke}, {Fortin}, {Frailis}, {Fukazawa}, {Funk}, {Fusco}, {Gargano}, {Gasparrini}, {Gehrels}, {Germani}, {Giebels}, {Giglietto}, {Giommi}, {Giordano}, {Glanzman}, {Godfrey}, {Grenier}, {Grondin}, {Grove}, {Guiriec}, {Hadasch}, {Hayashida}, {Hays}, {Healey}, {Horan}, {Hughes}, {Itoh}, {J{\'o}hannesson}, {Johnson}, {Johnson}, {Kamae}, {Katagiri}, {Kataoka}, {Kawai}, {Kn{\"o}dlseder}, {Kuss}, {Lande},
  {Larsson}, {Latronico}, {Lemoine-Goumard}, {Longo}, {Loparco}, {Lott}, {Lovellette}, {Lubrano}, {Madejski}, {Makeev}, {Massaro}, {Mazziotta}, {McEnery}, {Michelson}, {Mitthumsiri}, {Mizuno}, {Moiseev}, {Monte}, {Monzani}, {Morselli}, {Moskalenko}, {Mueller}, {Murgia}, {Nolan}, {Norris}, {Nuss}, {Ohno}, {Ohsugi}, {Omodei}, {Orlando}, {Ormes}, {Ozaki}, {Panetta}, {Parent}, {Pelassa}, {Pepe}, {Pesce-Rollins}, {Piron}, {Porter}, {Rain{\`o}}, {Rando}, {Razzano}, {Reimer}, {Reimer}, {Ritz}, {Rodriguez}, {Romani}, {Roth}, {Ryde}, {Sadrozinski}, {Sander}, {Scargle}, {Sgr{\`o}}, {Shaw}, {Smith}, {Spandre}, {Spinelli}, {Starck}, {Strickman}, {Suson}, {Takahashi}, {Takahashi}, {Tanaka}, {Thayer}, {Thayer}, {Thompson}, {Tibaldo}, {Torres}, {Tosti}, {Tramacere}, {Uchiyama}, {Usher}, {Vasileiou}, {Vilchez}, {Vitale}, {Waite}, {Wallace}, {Wang}, {Winer}, {Wood}, {Yang}, {Ylinen}, \& {Ziegler}}]{2010ApJ...722..520A}
{Abdo}, A.~A., {Ackermann}, M., {Ajello}, M., {et~al.} 2010{\natexlab{b}}, \apj, 722, 520

\bibitem[{{Abdo} {et~al.}(2010{\natexlab{c}}){Abdo}, {Ackermann}, {Ajello}, {Atwood}, {Axelsson}, {Baldini}, {Ballet}, {Barbiellini}, {Bastieri}, {Bechtol}, {Bellazzini}, {Berenji}, {Blandford}, {Bloom}, {Bonamente}, {Borgland}, {Bouvier}, {Bregeon}, {Brez}, {Brigida}, {Bruel}, {Burnett}, {Buson}, {Caliandro}, {Cameron}, {Caraveo}, {Carrigan}, {Casandjian}, {Cavazzuti}, {Cecchi}, {{\c{C}}elik}, {Charles}, {Chekhtman}, {Cheung}, {Chiang}, {Ciprini}, {Claus}, {Cohen-Tanugi}, {Conrad}, {Cutini}, {Dermer}, {de Angelis}, {de Palma}, {Digel}, {Silva}, {Drell}, {Dubois}, {Dumora}, {Farnier}, {Favuzzi}, {Fegan}, {Focke}, {Fortin}, {Frailis}, {Fukazawa}, {Funk}, {Fusco}, {Gargano}, {Gasparrini}, {Gehrels}, {Germani}, {Giebels}, {Giglietto}, {Giommi}, {Giordano}, {Glanzman}, {Godfrey}, {Grenier}, {Grondin}, {Grove}, {Guillemot}, {Guiriec}, {Harding}, {Hartman}, {Hayashida}, {Hays}, {Healey}, {Horan}, {Hughes}, {Jackson}, {J{\'o}hannesson}, {Johnson}, {Johnson}, {Kamae}, {Katagiri}, {Kataoka}, {Kawai}, {Kerr},
  {Kn{\"o}dlseder}, {Kuss}, {Lande}, {Latronico}, {Lemoine-Goumard}, {Longo}, {Loparco}, {Lott}, {Lovellette}, {Lubrano}, {Madejski}, {Makeev}, {Mazziotta}, {McConville}, {McEnery}, {Meurer}, {Michelson}, {Mitthumsiri}, {Mizuno}, {Moiseev}, {Monte}, {Monzani}, {Morselli}, {Moskalenko}, {Murgia}, {Nolan}, {Norris}, {Nuss}, {Ohsugi}, {Omodei}, {Orlando}, {Ormes}, {Paneque}, {Panetta}, {Parent}, {Pelassa}, {Pepe}, {Persic}, {Pesce-Rollins}, {Piron}, {Porter}, {Rain{\`o}}, {Rando}, {Razzano}, {Reimer}, {Reimer}, {Reposeur}, {Ritz}, {Rochester}, {Rodriguez}, {Romani}, {Roth}, {Ryde}, {Sadrozinski}, {Sanchez}, {Sander}, {Saz Parkinson}, {Scargle}, {Sgr{\`o}}, {Siskind}, {Smith}, {Smith}, {Spandre}, {Spinelli}, {Strickman}, {Suson}, {Tajima}, {Takahashi}, {Takahashi}, {Tanaka}, {Thayer}, {Thayer}, {Thompson}, {Tibaldo}, {Torres}, {Tosti}, {Tramacere}, {Uchiyama}, {Usher}, {Vasileiou}, {Vilchez}, {Villata}, {Vitale}, {Waite}, {Wang}, {Winer}, {Wood}, {Ylinen}, \& {Ziegler}}]{2010ApJ...710.1271A}
{Abdo}, A.~A., {Ackermann}, M., {Ajello}, M., {et~al.} 2010{\natexlab{c}}, \apj, 710, 1271

\bibitem[{{Abdo} {et~al.}(2011){Abdo}, {Ackermann}, {Ajello}, {Baldini}, {Ballet}, {Barbiellini}, {Bastieri}, {Bechtol}, {Bellazzini}, {Berenji}, {Blandford}, {Bonamente}, {Borgland}, {Bouvier}, {Bregeon}, {Brez}, {Brigida}, {Bruel}, {Buehler}, {Buson}, {Caliandro}, {Cameron}, {Caraveo}, {Carrigan}, {Casandjian}, {Cavazzuti}, {Cecchi}, {{\c{C}}elik}, {Charles}, {Chekhtman}, {Cheung}, {Chiang}, {Ciprini}, {Claus}, {Cohen-Tanugi}, {Conrad}, {Costamante}, {Cutini}, {Davis}, {Dermer}, {de Palma}, {Digel}, {do Couto e Silva}, {Drell}, {Dubois}, {Dumora}, {Favuzzi}, {Fegan}, {Fortin}, {Frailis}, {Fuhrmann}, {Fukazawa}, {Funk}, {Fusco}, {Gargano}, {Gasparrini}, {Gehrels}, {Germani}, {Giglietto}, {Giommi}, {Giordano}, {Giroletti}, {Glanzman}, {Godfrey}, {Grenier}, {Grove}, {Guillemot}, {Guiriec}, {Hadasch}, {Hayashida}, {Hays}, {Horan}, {Hughes}, {Itoh}, {J{\'o}hannesson}, {Johnson}, {Johnson}, {Johnson}, {Kamae}, {Katagiri}, {Kataoka}, {Kn{\"o}dlseder}, {Kuss}, {Lande}, {Latronico}, {Lee}, {Longo}, {Loparco},
  {Lott}, {Lovellette}, {Lubrano}, {Makeev}, {Mazziotta}, {McEnery}, {Mehault}, {Michelson}, {Mizuno}, {Moiseev}, {Monte}, {Monzani}, {Morselli}, {Moskalenko}, {Murgia}, {Nakamori}, {Naumann-Godo}, {Nestoras}, {Nolan}, {Norris}, {Nuss}, {Ohsugi}, {Okumura}, {Omodei}, {Orlando}, {Ormes}, {Ozaki}, {Paneque}, {Panetta}, {Parent}, {Pelassa}, {Pepe}, {Pesce-Rollins}, {Piron}, {Porter}, {Rain{\`o}}, {Rando}, {Razzano}, {Reimer}, {Reimer}, {Reyes}, {Ripken}, {Ritz}, {Romani}, {Roth}, {Sadrozinski}, {Sanchez}, {Sander}, {Scargle}, {Sgr{\`o}}, {Shaw}, {Smith}, {Spandre}, {Spinelli}, {Strickman}, {Suson}, {Takahashi}, {Tanaka}, {Thayer}, {Thayer}, {Thompson}, {Tibaldo}, {Torres}, {Tosti}, {Tramacere}, {Usher}, {Vandenbroucke}, {Vasileiou}, {Vilchez}, {Vitale}, {Waite}, {Wang}, {Winer}, {Wood}, {Yang}, {Ylinen}, {Ziegler}, {Acciari}, {Aliu}, {Arlen}, {Aune}, {Beilicke}, {Benbow}, {B{\"o}ttcher}, {Boltuch}, {Bradbury}, {Buckley}, {Bugaev}, {Byrum}, {Cannon}, {Cesarini}, {Christiansen}, {Ciupik}, {Cui}, {de la Calle
  Perez}, {Dickherber}, {Errando}, {Falcone}, {Finley}, {Finnegan}, {Fortson}, {Furniss}, {Galante}, {Gall}, {Gillanders}, {Godambe}, {Grube}, {Guenette}, {Gyuk}, {Hanna}, {Holder}, {Hui}, {Humensky}, {Imran}, {Kaaret}, {Karlsson}, {Kertzman}, {Kieda}, {Konopelko}, {Krawczynski}, {Krennrich}, {Lang}, {LeBohec}, {Maier}, {McArthur}, {McCann}, {McCutcheon}, {Moriarty}, {Mukherjee}, {Ong}, {Otte}, {Pandel}, {Perkins}, {Pichel}, {Pohl}, {Quinn}, {Ragan}, {Reynolds}, {Roache}, {Rose}, {Schroedter}, {Sembroski}, {Senturk}, {Smith}, {Steele}, {Swordy}, {Te{\v{s}}i{\'c}}, {Theiling}, {Thibadeau}, {Varlotta}, {Vassiliev}, {Vincent}, {Wakely}, {Ward}, {Weekes}, {Weinstein}, {Weisgarber}, {Williams}, {Wissel}, {Wood}, {Villata}, {Raiteri}, {Gurwell}, {Larionov}, {Kurtanidze}, {Aller}, {L{\"a}hteenm{\"a}ki}, {Chen}, {Berduygin}, {Agudo}, {Aller}, {Arkharov}, {Bach}, {Bachev}, {Beltrame}, {Ben{\'\i}tez}, {Buemi}, {Dashti}, {Calcidese}, {Capezzali}, {Carosati}, {Da Rio}, {Di Paola}, {Diltz}, {Dolci}, {Dultzin},
  {Forn{\'e}}, {G{\'o}mez}, {Hagen-Thorn}, {Halkola}, {Heidt}, {Hiriart}, {Hovatta}, {Hsiao}, {Jorstad}, {Kimeridze}, {Konstantinova}, {Kopatskaya}, {Koptelova}, {Leto}, {Ligustri}, {Lindfors}, {Lopez}, {Marscher}, {Mommert}, {Mujica}, {Nikolashvili}, {Nilsson}, {Palma}, {Pasanen}, {Roca-Sogorb}, {Ros}, {Roustazadeh}, {Sadun}, {Saino}, {Sigua}, {Sillan{\"a}{\"a}}, {Sorcia}, {Takalo}, {Tornikoski}, {Trigilio}, {Turchetti}, {Umana}, {Belloni}, {Blake}, {Bloom}, {Angelakis}, {Fumagalli}, {Hauser}, {Prochaska}, {Riquelme}, {Sievers}, {Starr}, {Tagliaferri}, {Ungerechts}, {Wagner}, {Zensus}, {Fermi LAT Collaboration}, {VERITAS Collaboration}, \& {GASP-WEBT Consortium}}]{2011ApJ...726...43A}
{Abdo}, A.~A., {Ackermann}, M., {Ajello}, M., {et~al.} 2011, \apj, 726, 43

\bibitem[{{Abdo} {et~al.}(2009){Abdo}, {Ackermann}, {Atwood}, {Axelsson}, {Baldini}, {Ballet}, {Barbiellini}, {Bastieri}, {Baughman}, {Bechtol}, {Bellazzini}, {Berenji}, {Blandford}, {Bloom}, {Bogaert}, {Bonamente}, {Borgland}, {Bregeon}, {Brez}, {Brigida}, {Bruel}, {Burnett}, {Caliandro}, {Cameron}, {Caraveo}, {Casandjian}, {Cavazzuti}, {Cecchi}, {Charles}, {Chekhtman}, {Cheung}, {Chiang}, {Ciprini}, {Claus}, {Cohen-Tanugi}, {Colafrancesco}, {Conrad}, {Costamante}, {Cutini}, {Dermer}, {de Angelis}, {de Palma}, {Digel}, {do Couto e Silva}, {Drell}, {Dubois}, {Dumora}, {Edmonds}, {Farnier}, {Favuzzi}, {Ferrara}, {Fleury}, {Focke}, {Foschini}, {Frailis}, {Fuhrmann}, {Fukazawa}, {Funk}, {Fusco}, {Gargano}, {Gasparrini}, {Gehrels}, {Germani}, {Giebels}, {Giglietto}, {Giordano}, {Giroletti}, {Glanzman}, {Godfrey}, {Grenier}, {Grondin}, {Grove}, {Guillemot}, {Guiriec}, {Harding}, {Hartman}, {Hayashida}, {Hays}, {Healey}, {Hughes}, {J{\'o}hannesson}, {Johnson}, {Johnson}, {Johnson}, {Kadler}, {Kamae}, {Katagiri},
  {Kataoka}, {Kawai}, {Kerr}, {Kn{\"o}dlseder}, {Kocian}, {Kuehn}, {Kuss}, {Latronico}, {Lee}, {Lemoine-Goumard}, {Longo}, {Loparco}, {Lott}, {Lovellette}, {Lubrano}, {Madejski}, {Makeev}, {Marelli}, {Mazziotta}, {McEnery}, {McGlynn}, {Meurer}, {Michelson}, {Mitthumsiri}, {Mizuno}, {Moiseev}, {Monte}, {Monzani}, {Morselli}, {Moskalenko}, {Murgia}, {Nolan}, {Nuss}, {Ohno}, {Ohsugi}, {Ojha}, {Omodei}, {Orlando}, {Ormes}, {Paneque}, {Panetta}, {Parent}, {Pepe}, {Pesce-Rollins}, {Piron}, {Porter}, {Rain{\`o}}, {Rando}, {Razzano}, {Reimer}, {Reimer}, {Reposeur}, {Reyes}, {Ritz}, {Rochester}, {Rodriguez}, {Romani}, {Roth}, {Ryde}, {Sadrozinski}, {Sambruna}, {Sanchez}, {Sander}, {Parkinson}, {Sgr{\`o}}, {Shaw}, {Siskind}, {Smith}, {Smith}, {Spandre}, {Spinelli}, {Starck}, {Strickman}, {Suson}, {Tajima}, {Takahashi}, {Tanaka}, {Thayer}, {Thayer}, {Thompson}, {Tibaldo}, {Tibolla}, {Torres}, {Tosti}, {Tramacere}, {Usher}, {Vilchez}, {Vitale}, {Waite}, {Wang}, {Winer}, {Wood}, {Ylinen}, {Ziegler}, {Fermi/LAT
  Collaboration}, {Edwards}, {Chester}, {Burrows}, {Hauser}, \& {Wagner}}]{2009ApJ...697..934A}
{Abdo}, A.~A., {Ackermann}, M., {Atwood}, W.~B., {et~al.} 2009, \apj, 697, 934

\bibitem[{{Abdollahi} {et~al.}(2020){Abdollahi}, {Acero}, {Ackermann}, {Ajello}, {Atwood}, {Axelsson}, {Baldini}, {Ballet}, {Barbiellini}, {Bastieri}, {Becerra Gonzalez}, {Bellazzini}, {Berretta}, {Bissaldi}, {Blandford}, {Bloom}, {Bonino}, {Bottacini}, {Brandt}, {Bregeon}, {Bruel}, {Buehler}, {Burnett}, {Buson}, {Cameron}, {Caputo}, {Caraveo}, {Casandjian}, {Castro}, {Cavazzuti}, {Charles}, {Chaty}, {Chen}, {Cheung}, {Chiaro}, {Ciprini}, {Cohen-Tanugi}, {Cominsky}, {Coronado-Bl{\'a}zquez}, {Costantin}, {Cuoco}, {Cutini}, {D'Ammando}, {DeKlotz}, {de la Torre Luque}, {de Palma}, {Desai}, {Digel}, {Di Lalla}, {Di Mauro}, {Di Venere}, {Dom{\'\i}nguez}, {Dumora}, {Fana Dirirsa}, {Fegan}, {Ferrara}, {Franckowiak}, {Fukazawa}, {Funk}, {Fusco}, {Gargano}, {Gasparrini}, {Giglietto}, {Giommi}, {Giordano}, {Giroletti}, {Glanzman}, {Green}, {Grenier}, {Griffin}, {Grondin}, {Grove}, {Guiriec}, {Harding}, {Hayashi}, {Hays}, {Hewitt}, {Horan}, {J{\'o}hannesson}, {Johnson}, {Kamae}, {Kerr}, {Kocevski}, {Kovac'evic'},
  {Kuss}, {Landriu}, {Larsson}, {Latronico}, {Lemoine-Goumard}, {Li}, {Liodakis}, {Longo}, {Loparco}, {Lott}, {Lovellette}, {Lubrano}, {Madejski}, {Maldera}, {Malyshev}, {Manfreda}, {Marchesini}, {Marcotulli}, {Mart{\'\i}-Devesa}, {Martin}, {Massaro}, {Mazziotta}, {McEnery}, {Mereu}, {Meyer}, {Michelson}, {Mirabal}, {Mizuno}, {Monzani}, {Morselli}, {Moskalenko}, {Negro}, {Nuss}, {Ojha}, {Omodei}, {Orienti}, {Orlando}, {Ormes}, {Palatiello}, {Paliya}, {Paneque}, {Pei}, {Pe{\~n}a-Herazo}, {Perkins}, {Persic}, {Pesce-Rollins}, {Petrosian}, {Petrov}, {Piron}, {Poon}, {Porter}, {Principe}, {Rain{\`o}}, {Rando}, {Razzano}, {Razzaque}, {Reimer}, {Reimer}, {Remy}, {Reposeur}, {Romani}, {Saz Parkinson}, {Schinzel}, {Serini}, {Sgr{\`o}}, {Siskind}, {Smith}, {Spandre}, {Spinelli}, {Strong}, {Suson}, {Tajima}, {Takahashi}, {Tak}, {Thayer}, {Thompson}, {Tibaldo}, {Torres}, {Torresi}, {Valverde}, {Van Klaveren}, {van Zyl}, {Wood}, {Yassine}, \& {Zaharijas}}]{2020ApJS..247...33A}
{Abdollahi}, S., {Acero}, F., {Ackermann}, M., {et~al.} 2020, \apjs, 247, 33

\bibitem[{{Abramowicz} {et~al.}(1980){Abramowicz}, {Calvani}, \& {Nobili}}]{1980ApJ...242..772A}
{Abramowicz}, M.~A., {Calvani}, M., \& {Nobili}, L. 1980, \apj, 242, 772

\bibitem[{{Acciari} {et~al.}(2008){Acciari}, {Aliu}, {Beilicke}, {Benbow}, {B{\"o}ttcher}, {Bradbury}, {Buckley}, {Bugaev}, {Butt}, {Celik}, {Cesarini}, {Ciupik}, {Chow}, {Cogan}, {Colin}, {Cui}, {Daniel}, {Ergin}, {Falcone}, {Fegan}, {Finley}, {Finnegan}, {Fortin}, {Fortson}, {Furniss}, {Gall}, {Gillanders}, {Grube}, {Guenette}, {Gyuk}, {Hanna}, {Hays}, {Holder}, {Horan}, {Hui}, {Humensky}, {Imran}, {Kaaret}, {Karlsson}, {Kertzman}, {Kieda}, {Konopelko}, {Krawczynski}, {Krennrich}, {Lang}, {LeBohec}, {Lee}, {Maier}, {McCann}, {McCutcheon}, {Moriarty}, {Mukherjee}, {Nagai}, {Niemiec}, {Ong}, {Pandel}, {Perkins}, {Petry}, {Pohl}, {Quinn}, {Ragan}, {Reyes}, {Reynolds}, {Roache}, {Rose}, {Schroedter}, {Sembroski}, {Smith}, {Steele}, {Swordy}, {Toner}, {Vassiliev}, {Wagner}, {Wakely}, {Ward}, {Weekes}, {Weinstein}, {White}, {Williams}, {Wissel}, {Wood}, \& {Zitzer}}]{2008ApJ...684L..73A}
{Acciari}, V.~A., {Aliu}, E., {Beilicke}, M., {et~al.} 2008, \apjl, 684, L73

\bibitem[{{Ackermann} {et~al.}(2015){Ackermann}, {Ajello}, {Atwood}, {Baldini}, {Ballet}, {Barbiellini}, {Bastieri}, {Becerra Gonzalez}, {Bellazzini}, {Bissaldi}, {Blandford}, {Bloom}, {Bonino}, {Bottacini}, {Brandt}, {Bregeon}, {Britto}, {Bruel}, {Buehler}, {Buson}, {Caliandro}, {Cameron}, {Caragiulo}, {Caraveo}, {Carpenter}, {Casandjian}, {Cavazzuti}, {Cecchi}, {Charles}, {Chekhtman}, {Cheung}, {Chiang}, {Chiaro}, {Ciprini}, {Claus}, {Cohen-Tanugi}, {Cominsky}, {Conrad}, {Cutini}, {D'Abrusco}, {D'Ammando}, {de Angelis}, {Desiante}, {Digel}, {Di Venere}, {Drell}, {Favuzzi}, {Fegan}, {Ferrara}, {Finke}, {Focke}, {Franckowiak}, {Fuhrmann}, {Fukazawa}, {Furniss}, {Fusco}, {Gargano}, {Gasparrini}, {Giglietto}, {Giommi}, {Giordano}, {Giroletti}, {Glanzman}, {Godfrey}, {Grenier}, {Grove}, {Guiriec}, {Hewitt}, {Hill}, {Horan}, {Itoh}, {J{\'o}hannesson}, {Johnson}, {Johnson}, {Kataoka}, {Kawano}, {Krauss}, {Kuss}, {La Mura}, {Larsson}, {Latronico}, {Leto}, {Li}, {Li}, {Longo}, {Loparco}, {Lott}, {Lovellette},
  {Lubrano}, {Madejski}, {Mayer}, {Mazziotta}, {McEnery}, {Michelson}, {Mizuno}, {Moiseev}, {Monzani}, {Morselli}, {Moskalenko}, {Murgia}, {Nuss}, {Ohno}, {Ohsugi}, {Ojha}, {Omodei}, {Orienti}, {Orlando}, {Paggi}, {Paneque}, {Perkins}, {Pesce-Rollins}, {Piron}, {Pivato}, {Porter}, {Rain{\`o}}, {Rando}, {Razzano}, {Razzaque}, {Reimer}, {Reimer}, {Romani}, {Salvetti}, {Schaal}, {Schinzel}, {Schulz}, {Sgr{\`o}}, {Siskind}, {Sokolovsky}, {Spada}, {Spandre}, {Spinelli}, {Stawarz}, {Suson}, {Takahashi}, {Takahashi}, {Tanaka}, {Thayer}, {Thayer}, {Tibaldo}, {Torres}, {Torresi}, {Tosti}, {Troja}, {Uchiyama}, {Vianello}, {Winer}, {Wood}, \& {Zimmer}}]{2015ApJ...810...14A}
{Ackermann}, M., {Ajello}, M., {Atwood}, W.~B., {et~al.} 2015, \apj, 810, 14

\bibitem[{{Ag{\'\i}s-Gonz{\'a}lez} {et~al.}(2014){Ag{\'\i}s-Gonz{\'a}lez}, {Miniutti}, {Kara}, {Fabian}, {Sanfrutos}, {Risaliti}, {Bianchi}, {Strotjohann}, {Saxton}, \& {Parker}}]{2014MNRAS.443.2862A}
{Ag{\'\i}s-Gonz{\'a}lez}, B., {Miniutti}, G., {Kara}, E., {et~al.} 2014, \mnras, 443, 2862

\bibitem[{{Ai} {et~al.}(2020){Ai}, {Dou}, {Yang}, {Sun}, {Xie}, {Yao}, {Wu}, {Wang}, {Shu}, \& {Jiang}}]{2020ApJ...890L..29A}
{Ai}, Y., {Dou}, L., {Yang}, C., {et~al.} 2020, \apjl, 890, L29

\bibitem[{{Aleksi{\'c}} {et~al.}(2014){Aleksi{\'c}}, {Ansoldi}, {Antonelli}, {Antoranz}, {Babic}, {Bangale}, {Barres de Almeida}, {Barrio}, {Becerra Gonz{\'a}lez}, {Bednarek}, {Berger}, {Bernardini}, {Biland}, {Blanch}, {Bock}, {Bonnefoy}, {Bonnoli}, {Borracci}, {Bretz}, {Carmona}, {Carosi}, {Carreto Fidalgo}, {Colin}, {Colombo}, {Contreras}, {Cortina}, {Covino}, {Da Vela}, {Dazzi}, {De Angelis}, {De Caneva}, {De Lotto}, {Delgado Mendez}, {Doert}, {Dom{\'\i}nguez}, {Dominis Prester}, {Dorner}, {Doro}, {Einecke}, {Eisenacher}, {Elsaesser}, {Farina}, {Ferenc}, {Fonseca}, {Font}, {Frantzen}, {Fruck}, {Garc{\'\i}a L{\'o}pez}, {Garczarczyk}, {Garrido Terrats}, {Gaug}, {Giavitto}, {Godinovi{\'c}}, {Gonz{\'a}lez Mu{\~n}oz}, {Gozzini}, {Hadasch}, {Hayashida}, {Herrero}, {Hildebrand}, {Hose}, {Hrupec}, {Idec}, {Kadenius}, {Kellermann}, {Kodani}, {Konno}, {Krause}, {Kubo}, {Kushida}, {La Barbera}, {Lelas}, {Lewandowska}, {Lindfors}, {Lombardi}, {L{\'o}pez}, {L{\'o}pez-Coto}, {L{\'o}pez-Oramas}, {Lorenz}, {Lozano},
  {Makariev}, {Mallot}, {Maneva}, {Mankuzhiyil}, {Mannheim}, {Maraschi}, {Marcote}, {Mariotti}, {Mart{\'\i}nez}, {Mazin}, {Menzel}, {Meucci}, {Miranda}, {Mirzoyan}, {Moralejo}, {Munar-Adrover}, {Nakajima}, {Niedzwiecki}, {Nilsson}, {Nishijima}, {Nowak}, {Orito}, {Overkemping}, {Paiano}, {Palatiello}, {Paneque}, {Paoletti}, {Paredes}, {Paredes-Fortuny}, {Partini}, {Persic}, {Prada}, {Prada Moroni}, {Prandini}, {Preziuso}, {Puljak}, {Reinthal}, {Rhode}, {Rib{\'o}}, {Rico}, {Rodriguez Garcia}, {R{\"u}gamer}, {Saggion}, {Saito}, {Saito}, {Salvati}, {Satalecka}, {Scalzotto}, {Scapin}, {Schultz}, {Schweizer}, {Shore}, {Sillanp{\"a}{\"a}}, {Sitarek}, {Snidaric}, {Sobczynska}, {Spanier}, {Stamatescu}, {Stamerra}, {Steinbring}, {Storz}, {Sun}, {Suri{\'c}}, {Takalo}, {Takami}, {Tavecchio}, {Temnikov}, {Terzi{\'c}}, {Tescaro}, {Teshima}, {Thaele}, {Tibolla}, {Torres}, {Toyama}, {Treves}, {Uellenbeck}, {Vogler}, {Wagner}, {Zandanel}, {Zanin}, {MAGIC Collaboration}, {Cutini}, {Gasparrini}, {Furniss}, {Hovatta}, {Kangas},
  {Kankare}, {Kotilainen}, {Lister}, {L{\"a}hteenm{\"a}ki}, {Max-Moerbeck}, {Pavlidou}, {Readhead}, \& {Richards}}]{2014A&A...567A.135A}
{Aleksi{\'c}}, J., {Ansoldi}, S., {Antonelli}, L.~A., {et~al.} 2014, \aap, 567, A135

\bibitem[{{{\'A}lvarez Crespo} {et~al.}(2016){{\'A}lvarez Crespo}, {Masetti}, {Ricci}, {Landoni}, {Pati{\~n}o-{\'A}lvarez}, {Massaro}, {D'Abrusco}, {Paggi}, {Chavushyan}, {Jim{\'e}nez-Bail{\'o}n}, {Torrealba}, {Latronico}, {La Franca}, {Smith}, \& {Tosti}}]{2016AJ....151...32A}
{{\'A}lvarez Crespo}, N., {Masetti}, N., {Ricci}, F., {et~al.} 2016, \aj, 151, 32

\bibitem[{{Aretxaga} {et~al.}(1999){Aretxaga}, {Joguet}, {Kunth}, {Melnick}, \& {Terlevich}}]{1999ApJ...519L.123A}
{Aretxaga}, I., {Joguet}, B., {Kunth}, D., {Melnick}, J., \& {Terlevich}, R.~J. 1999, \apjl, 519, L123

\bibitem[{{Bednarek}(1998)}]{1998MNRAS.294..439B}
{Bednarek}, W. 1998, \mnras, 294, 439

\bibitem[{{Blandford} \& {Levinson}(1995)}]{Blandford95}
{Blandford}, R.~D. \& {Levinson}, A. 1995, \apj, 441, 79

\bibitem[{{B{\"o}ttcher} \& {Chiang}(2002)}]{2002ApJ...581..127B}
{B{\"o}ttcher}, M. \& {Chiang}, J. 2002, \apj, 581, 127

\bibitem[{{B{\"o}ttcher} \& {Dermer}(2002)}]{2002ApJ...564...86B}
{B{\"o}ttcher}, M. \& {Dermer}, C.~D. 2002, \apj, 564, 86

\bibitem[{{Boula} {et~al.}(2019){Boula}, {Kazanas}, \& {Mastichiadis}}]{2019MNRAS.482L..80B}
{Boula}, S., {Kazanas}, D., \& {Mastichiadis}, A. 2019, \mnras, 482, L80

\bibitem[{{Burrows} {et~al.}(2005){Burrows}, {Hill}, {Nousek}, {Kennea}, {Wells}, {Osborne}, {Abbey}, {Beardmore}, {Mukerjee}, {Short}, {Chincarini}, {Campana}, {Citterio}, {Moretti}, {Pagani}, {Tagliaferri}, {Giommi}, {Capalbi}, {Tamburelli}, {Angelini}, {Cusumano}, {Br{\"a}uninger}, {Burkert}, \& {Hartner}}]{2005SSRv..120..165B}
{Burrows}, D.~N., {Hill}, J.~E., {Nousek}, J.~A., {et~al.} 2005, \ssr, 120, 165

\bibitem[{{Cao} \& {Wang}(2013)}]{2013MNRAS.436.2170C}
{Cao}, G. \& {Wang}, J.-C. 2013, \mnras, 436, 2170

\bibitem[{{Cao}(2002)}]{2002ApJ...570L..13C}
{Cao}, X. 2002, \apjl, 570, L13

\bibitem[{{Cao}(2003)}]{2003ApJ...599..147C}
{Cao}, X. 2003, \apj, 599, 147

\bibitem[{{Cao}(2010)}]{2010ApJ...724..855C}
{Cao}, X. 2010, \apj, 724, 855

\bibitem[{{Capelo} \& {Natarajan}(2007)}]{2007NJPh....9..445C}
{Capelo}, P.~R. \& {Natarajan}, P. 2007, New Journal of Physics, 9, 445

\bibitem[{{Cavaliere} \& {D'Elia}(2002)}]{2002ApJ...571..226C}
{Cavaliere}, A. \& {D'Elia}, V. 2002, \apj, 571, 226

\bibitem[{{Chen}(2014)}]{2014ApJ...788..179C}
{Chen}, L. 2014, \apj, 788, 179

\bibitem[{{Chen} \& {Bai}(2011)}]{2011ApJ...735..108C}
{Chen}, L. \& {Bai}, J.~M. 2011, \apj, 735, 108

\bibitem[{{Ciprini}(2018)}]{2018ATel12277....1C}
{Ciprini}, S. 2018, The Astronomer's Telegram, 12277, 1

\bibitem[{{Ciprini} \& {Cheung}(2020)}]{2020ATel13382....1C}
{Ciprini}, S. \& {Cheung}, C.~C. 2020, The Astronomer's Telegram, 13382, 1

\bibitem[{{D'Ammando} {et~al.}(2022){D'Ammando}, {Angioni}, {Orienti}, {The Fermi Large Area Telescope Collaboration}, {Sitarek}, {Nozaki}, {Lindfors}, {Bonnoli}, {Fallah Ramazani}, {Jorstad}, {The MAGIC Collaboration}, {Acciari}, {Ansoldi}, {Antonelli}, {Arbet Engels}, {Artero}, {Asano}, {Baack}, {Babic}, {Baquero}, {Barres de Almeida}, {Barrio}, {Batkovi{\'c}}, {Becerra Gonzalez}, {Bednarek}, {Bellizzi}, {Bernardini}, {Bernardos}, {Berti}, {Besenrieder}, {Bhattacharyya}, {Bigongiari}, {Biland}, {Blanch}, {B{\"o}kenkamp}, {Bosnjak}, {Busetto}, {Carosi}, {Ceribella}, {Cerruti}, {Chai}, {Chilingarian}, {Cikota}, {Colak}, {Colombo}, {Contreras}, {Cortina}, {Covino}, {D'Amico}, {D'Elia}, {da Vela}, {Dazzi}, {de Angelis}, {de Lotto}, {Delfino}, {Delgado}, {Delgado Mendez}, {Depaoli}, {di Pierro}, {di Venere}, {Do Souto Espi{\~n}eira}, {Dominis Prester}, {Donini}, {Dorner}, {Doro}, {Elsaesser}, {Fattorini}, {Fonseca}, {Font}, {Fruck}, {Fukami}, {Fukazawa}, {Garc{\'\i}a L{\'o}pez}, {Garczarczyk}, {Gasparyan},
  {Gaug}, {Giglietto}, {Giordano}, {Gliwny}, {Godinovic}, {Green}, {Green}, {Hadasch}, {Hahn}, {Heckmann}, {Herrera}, {Hoang}, {Hrupec}, {H{\"u}tten}, {Inada}, {Ishio}, {Iwamura}, {Jim{\'e}nez Mart{\'\i}nez}, {Jormanainen}, {Jouvin}, {Karjalainen}, {Kerszberg}, {Kobayashi}, {Kubo}, {Kushida}, {Lamastra}, {Lelas}, {Leone}, {Linhoff}, {Lombardi}, {Longo}, {Lopez-Coto}, {L{\'o}pez-Moya}, {L{\'o}pez-Oramas}, {Loporchio}, {Machado de Oliveira Fraga}, {Maggio}, {Majumdar}, {Makariev}, {Mallamaci}, {Maneva}, {Manganaro}, {Mannheim}, {Maraschi}, {Mariotti}, {Martinez}, {Mazin}, {Menchiari}, {Mender}, {Mi{\'c}anovi{\'c}}, {Miceli}, {Miener}, {Miranda}, {Mirzoyan}, {Molina}, {Moralejo}, {Morcuende}, {Moreno}, {Moretti}, {Nakamori}, {Nava}, {Neustroev}, {Nigro}, {Nilsson}, {Nishijima}, {Noda}, {Ohtani}, {Oka}, {Otero-Santos}, {Paiano}, {Palatiello}, {Paneque}, {Paoletti}, {Paredes}, {Pavleti{\'c}}, {Pe{\~n}il}, {Persic}, {Pihet}, {Prada Moroni}, {Prandini}, {Priyadarshi}, {Puljak}, {Rhode}, {Rib{\'o}}, {Rico}, {Righi},
  {Rugliancich}, {Sahakyan}, {Saito}, {Sakurai}, {Satalecka}, {Saturni}, {Schleicher}, {Schmidt}, {Schweizer}, {{\v{S}}nidari{\'c}}, {Sobczy{\'n}ska}, {Spolon}, {Stamerra}, {Stri{\v{s}}kovi{\'c}}, {Strom}, {Strzys}, {Suda}, {Suri{\'c}}, {Takahashi}, {Takeishi}, {Tavecchio}, {Temnikov}, {Terzic}, {Teshima}, {Tosti}, {Truzzi}, {Tutone}, {Ubach}, {van Scherpenberg}, {Vanzo}, {Vazquez Acosta}, {Ventura}, {Verguilov}, {Vigorito}, {Vitale}, {Vovk}, {Will}, {Wunderlich}, {Yamamoto}, \& {Zari{\'c}}}]{2022icrc.confE.775D}
{D'Ammando}, F., {Angioni}, R., {Orienti}, M., {et~al.} 2022, in 37th International Cosmic Ray Conference, 775

\bibitem[{{Danforth} {et~al.}(2016){Danforth}, {Stocke}, {France}, {Begelman}, \& {Perlman}}]{2016ApJ...832...76D}
{Danforth}, C.~W., {Stocke}, J.~T., {France}, K., {Begelman}, M.~C., \& {Perlman}, E. 2016, \apj, 832, 76

\bibitem[{{Denney} {et~al.}(2014){Denney}, {De Rosa}, {Croxall}, {Gupta}, {Bentz}, {Fausnaugh}, {Grier}, {Martini}, {Mathur}, {Peterson}, {Pogge}, \& {Shappee}}]{2014ApJ...796..134D}
{Denney}, K.~D., {De Rosa}, G., {Croxall}, K., {et~al.} 2014, \apj, 796, 134

\bibitem[{{Dermer} {et~al.}(2009){Dermer}, {Finke}, {Krug}, \& {B{\"o}ttcher}}]{2009ApJ...692...32D}
{Dermer}, C.~D., {Finke}, J.~D., {Krug}, H., \& {B{\"o}ttcher}, M. 2009, \apj, 692, 32

\bibitem[{{Dermer} {et~al.}(1992){Dermer}, {Schlickeiser}, \& {Mastichiadis}}]{Dermer92}
{Dermer}, C.~D., {Schlickeiser}, R., \& {Mastichiadis}, A. 1992, \aap, 256, L27

\bibitem[{{Donato} {et~al.}(2001){Donato}, {Ghisellini}, {Tagliaferri}, \& {Fossati}}]{2001A&A...375..739D}
{Donato}, D., {Ghisellini}, G., {Tagliaferri}, G., \& {Fossati}, G. 2001, \aap, 375, 739

\bibitem[{{Edelson} {et~al.}(2002){Edelson}, {Turner}, {Pounds}, {Vaughan}, {Markowitz}, {Marshall}, {Dobbie}, \& {Warwick}}]{2002ApJ...568..610E}
{Edelson}, R., {Turner}, T.~J., {Pounds}, K., {et~al.} 2002, \apj, 568, 610

\bibitem[{{Elitzur} \& {Shlosman}(2006)}]{2006ApJ...648L.101E}
{Elitzur}, M. \& {Shlosman}, I. 2006, \apjl, 648, L101

\bibitem[{{Fan} {et~al.}(2013){Fan}, {Yang}, {Zhang}, {Hua}, {Liu}, {Qin}, \& {Huang}}]{2013PASJ...65...25F}
{Fan}, J., {Yang}, J.~H., {Zhang}, J.-Y., {et~al.} 2013, \pasj, 65, 25

\bibitem[{{Fan} {et~al.}(2002){Fan}, {Cheng}, \& {Zhang}}]{2002PASJ...54..533F}
{Fan}, J.-H., {Cheng}, K.~S., \& {Zhang}, L. 2002, \pasj, 54, 533

\bibitem[{{Fan} {et~al.}(2015){Fan}, {Yang}, {Liu}, {Cai}, \& {Lin}}]{2015IJMPA..3045020F}
{Fan}, J.~H., {Yang}, J.~H., {Liu}, Y., {Cai}, W., \& {Lin}, C. 2015, International Journal of Modern Physics A, 30, 1545020

\bibitem[{{Feigelson} \& {Babu}(2012)}]{2012msma.book.....F}
{Feigelson}, E.~D. \& {Babu}, G.~J. 2012, {Modern Statistical Methods for Astronomy}

\bibitem[{{Feng} {et~al.}(2021){Feng}, {Hu}, {Li}, {Liu}, {Bai}, {Xing}, {Wang}, {Yang}, {Xiao}, \& {Lu}}]{2021ApJ...909...18F}
{Feng}, H.-C., {Hu}, C., {Li}, S.-S., {et~al.} 2021, \apj, 909, 18

\bibitem[{{Finke} {et~al.}(2008){Finke}, {Dermer}, \& {B{\"o}ttcher}}]{2008ApJ...686..181F}
{Finke}, J.~D., {Dermer}, C.~D., \& {B{\"o}ttcher}, M. 2008, \apj, 686, 181

\bibitem[{{Fossati} {et~al.}(2008){Fossati}, {Buckley}, {Bond}, {Bradbury}, {Carter-Lewis}, {Chow}, {Cui}, {Falcone}, {Finley}, {Gaidos}, {Grube}, {Holder}, {Horan}, {Horns}, {Jordan}, {Kieda}, {Kildea}, {Krawczynski}, {Krennrich}, {Lang}, {LeBohec}, {Lee}, {Moriarty}, {Ong}, {Petry}, {Quinn}, {Sembroski}, {Wakely}, \& {Weekes}}]{2008ApJ...677..906F}
{Fossati}, G., {Buckley}, J.~H., {Bond}, I.~H., {et~al.} 2008, \apj, 677, 906

\bibitem[{{Gammie} {et~al.}(1999){Gammie}, {Narayan}, \& {Blandford}}]{1999ApJ...516..177G}
{Gammie}, C.~F., {Narayan}, R., \& {Blandford}, R. 1999, \apj, 516, 177

\bibitem[{{Ghisellini}(2006)}]{2006smqw.confE..27G}
{Ghisellini}, G. 2006, in VI Microquasar Workshop: Microquasars and Beyond, 27.1

\bibitem[{{Ghisellini} {et~al.}(1998{\natexlab{a}}){Ghisellini}, {Celotti}, {Fossati}, {Maraschi}, \& {Comastri}}]{1998MNRAS.301..451G}
{Ghisellini}, G., {Celotti}, A., {Fossati}, G., {Maraschi}, L., \& {Comastri}, A. 1998{\natexlab{a}}, \mnras, 301, 451

\bibitem[{{Ghisellini} {et~al.}(1998{\natexlab{b}}){Ghisellini}, {Haardt}, \& {Svensson}}]{1998MNRAS.297..348G}
{Ghisellini}, G., {Haardt}, F., \& {Svensson}, R. 1998{\natexlab{b}}, \mnras, 297, 348

\bibitem[{{Ghisellini} \& {Madau}(1996)}]{1996MNRAS.280...67G}
{Ghisellini}, G. \& {Madau}, P. 1996, \mnras, 280, 67

\bibitem[{{Ghisellini} {et~al.}(2009){Ghisellini}, {Maraschi}, \& {Tavecchio}}]{2009MNRAS.396L.105G}
{Ghisellini}, G., {Maraschi}, L., \& {Tavecchio}, F. 2009, \mnras, 396, L105

\bibitem[{{Ghisellini} \& {Tavecchio}(2008)}]{2008MNRAS.387.1669G}
{Ghisellini}, G. \& {Tavecchio}, F. 2008, \mnras, 387, 1669

\bibitem[{{Ghisellini} \& {Tavecchio}(2009)}]{2009MNRAS.397..985G}
{Ghisellini}, G. \& {Tavecchio}, F. 2009, \mnras, 397, 985

\bibitem[{{Ghisellini} {et~al.}(2011){Ghisellini}, {Tavecchio}, {Foschini}, \& {Ghirlanda}}]{2011MNRAS.414.2674G}
{Ghisellini}, G., {Tavecchio}, F., {Foschini}, L., \& {Ghirlanda}, G. 2011, \mnras, 414, 2674

\bibitem[{{Ghisellini} {et~al.}(2010){Ghisellini}, {Tavecchio}, {Foschini}, {Ghirlanda}, {Maraschi}, \& {Celotti}}]{2010MNRAS.402..497G}
{Ghisellini}, G., {Tavecchio}, F., {Foschini}, L., {et~al.} 2010, \mnras, 402, 497

\bibitem[{{Giommi} {et~al.}(2012){Giommi}, {Padovani}, {Polenta}, {Turriziani}, {D'Elia}, \& {Piranomonte}}]{2012MNRAS.420.2899G}
{Giommi}, P., {Padovani}, P., {Polenta}, G., {et~al.} 2012, \mnras, 420, 2899

\bibitem[{{Graham} {et~al.}(2020){Graham}, {Ross}, {Stern}, {Drake}, {McKernan}, {Ford}, {Djorgovski}, {Mahabal}, {Glikman}, {Larson}, \& {Christensen}}]{2020MNRAS.491.4925G}
{Graham}, M.~J., {Ross}, N.~P., {Stern}, D., {et~al.} 2020, \mnras, 491, 4925

\bibitem[{{Hewett} \& {Wild}(2010)}]{2010MNRAS.405.2302H}
{Hewett}, P.~C. \& {Wild}, V. 2010, \mnras, 405, 2302

\bibitem[{{Hovatta} \& {Lindfors}(2019)}]{2019NewAR..8701541H}
{Hovatta}, T. \& {Lindfors}, E. 2019, \nar, 87, 101541

\bibitem[{{Ivezi{\'c}} {et~al.}(2020){Ivezi{\'c}}, {Connolly}, {VanderPlas}, \& {Gray}}]{2020sdmm.book.....I}
{Ivezi{\'c}}, {\v{Z}}., {Connolly}, A.~J., {VanderPlas}, J.~T., \& {Gray}, A. 2020, {Statistics, Data Mining, and Machine Learning in Astronomy. A Practical Python Guide for the Analysis of Survey Data, Updated Edition}

\bibitem[{{Kang}(2017)}]{2017ApJ...837...38K}
{Kang}, S.-J. 2017, \apj, 837, 38

\bibitem[{{Kang} {et~al.}(2014){Kang}, {Chen}, \& {Wu}}]{2014ApJS..215....5K}
{Kang}, S.-J., {Chen}, L., \& {Wu}, Q. 2014, \apjs, 215, 5

\bibitem[{{Kang} \& {Lyu}(2022)}]{2022ATel15793....1K}
{Kang}, S.~J. \& {Lyu}, B. 2022, The Astronomer's Telegram, 15793, 1

\bibitem[{{Kang} {et~al.}(2024){Kang}, {Lyu}, {Wu}, {Zheng}, \& {Fan}}]{2024ApJ...962..122K}
{Kang}, S.-J., {Lyu}, B., {Wu}, Q., {Zheng}, Y.-G., \& {Fan}, J. 2024, \apj, 962, 122

\bibitem[{{Kang} {et~al.}(2023){Kang}, {Zheng}, \& {Wu}}]{2023MNRAS.525.3201K}
{Kang}, S.-J., {Zheng}, Y.-G., \& {Wu}, Q. 2023, \mnras, 525, 3201

\bibitem[{{Kang} {et~al.}(2016){Kang}, {Zheng}, {Wu}, \& {Chen}}]{2016MNRAS.461.1862K}
{Kang}, S.-J., {Zheng}, Y.-G., {Wu}, Q., \& {Chen}, L. 2016, \mnras, 461, 1862

\bibitem[{{Kang} {et~al.}(2021){Kang}, {Zheng}, {Wu}, {Chen}, \& {Yin}}]{2021MNRAS.502.5875K}
{Kang}, S.-J., {Zheng}, Y.-G., {Wu}, Q., {Chen}, L., \& {Yin}, Y. 2021, \mnras, 502, 5875

\bibitem[{{Kollatschny} {et~al.}(2020){Kollatschny}, {Grupe}, {Parker}, {Ochmann}, {Schartel}, {Herwig}, {Komossa}, {Romero-Colmenero}, \& {Santos-Lleo}}]{2020A&A...638A..91K}
{Kollatschny}, W., {Grupe}, D., {Parker}, M.~L., {et~al.} 2020, \aap, 638, A91

\bibitem[{{Kollatschny} {et~al.}(2018){Kollatschny}, {Ochmann}, {Zetzl}, {Haas}, {Chelouche}, {Kaspi}, {Pozo Nu{\~n}ez}, \& {Grupe}}]{2018A&A...619A.168K}
{Kollatschny}, W., {Ochmann}, M.~W., {Zetzl}, M., {et~al.} 2018, \aap, 619, A168

\bibitem[{{Krawczynski} {et~al.}(2004){Krawczynski}, {Hughes}, {Horan}, {Aharonian}, {Aller}, {Aller}, {Boltwood}, {Buckley}, {Coppi}, {Fossati}, {G{\"o}tting}, {Holder}, {Horns}, {Kurtanidze}, {Marscher}, {Nikolashvili}, {Remillard}, {Sadun}, \& {Schr{\"o}der}}]{2004ApJ...601..151K}
{Krawczynski}, H., {Hughes}, S.~B., {Horan}, D., {et~al.} 2004, \apj, 601, 151

\bibitem[{{LaMassa} {et~al.}(2015){LaMassa}, {Cales}, {Moran}, {Myers}, {Richards}, {Eracleous}, {Heckman}, {Gallo}, \& {Urry}}]{2015ApJ...800..144L}
{LaMassa}, S.~M., {Cales}, S., {Moran}, E.~C., {et~al.} 2015, \apj, 800, 144

\bibitem[{{LaMassa} {et~al.}(2017){LaMassa}, {Yaqoob}, \& {Kilgard}}]{2017ApJ...840...11L}
{LaMassa}, S.~M., {Yaqoob}, T., \& {Kilgard}, R. 2017, \apj, 840, 11

\bibitem[{{Lasota} {et~al.}(1996){Lasota}, {Abramowicz}, {Chen}, {Krolik}, {Narayan}, \& {Yi}}]{1996ApJ...462..142L}
{Lasota}, J.~P., {Abramowicz}, M.~A., {Chen}, X., {et~al.} 1996, \apj, 462, 142

\bibitem[{{Liao} {et~al.}(2014){Liao}, {Bai}, {Liu}, {Weng}, {Chen}, \& {Li}}]{2014ApJ...783...83L}
{Liao}, N.~H., {Bai}, J.~M., {Liu}, H.~T., {et~al.} 2014, \apj, 783, 83

\bibitem[{{Liao} {et~al.}(2015){Liao}, {Bai}, {Wang}, {Liu}, {Zhang}, {Jiang}, {Yuan}, \& {Chen}}]{2015RAA....15..313L}
{Liao}, N.-H., {Bai}, J.-M., {Wang}, J.-G., {et~al.} 2015, Research in Astronomy and Astrophysics, 15, 313

\bibitem[{{Lin} \& {Shields}(1986)}]{1986ApJ...305...28L}
{Lin}, D.~N.~C. \& {Shields}, G.~A. 1986, \apj, 305, 28

\bibitem[{{Liu} \& {Bai}(2006)}]{2006ApJ...653.1089L}
{Liu}, H.~T. \& {Bai}, J.~M. 2006, \apj, 653, 1089

\bibitem[{{Liu} {et~al.}(2008){Liu}, {Bai}, \& {Ma}}]{2008ApJ...688..148L}
{Liu}, H.~T., {Bai}, J.~M., \& {Ma}, L. 2008, \apj, 688, 148

\bibitem[{{Lyu} {et~al.}(2022){Lyu}, {Wu}, {Yan}, {Yu}, \& {Liu}}]{2022ApJ...927..227L}
{Lyu}, B., {Wu}, Q., {Yan}, Z., {Yu}, W., \& {Liu}, H. 2022, \apj, 927, 227

\bibitem[{{MAGIC Collaboration} {et~al.}(2021){MAGIC Collaboration}, {Acciari}, {Ansoldi}, {Antonelli}, {Arbet Engels}, {Artero}, {Asano}, {Baack}, {Babi{\'c}}, {Baquero}, {Barres de Almeida}, {Barrio}, {Becerra Gonz{\'a}lez}, {Bednarek}, {Bellizzi}, {Bernardini}, {Bernardos}, {Berti}, {Besenrieder}, {Bhattacharyya}, {Bigongiari}, {Biland}, {Blanch}, {Bonnoli}, {Bo{\v{s}}njak}, {Busetto}, {Carosi}, {Ceribella}, {Cerruti}, {Chai}, {Chilingarian}, {Cikota}, {Colak}, {Colombo}, {Contreras}, {Cortina}, {Covino}, {D'Amico}, {D'Elia}, {da Vela}, {Dazzi}, {de Angelis}, {de Lotto}, {Delfino}, {Delgado}, {Delgado Mendez}, {Depaoli}, {di Pierro}, {di Venere}, {Do Souto Espi{\~n}eira}, {Dominis Prester}, {Donini}, {Dorner}, {Doro}, {Elsaesser}, {Fallah Ramazani}, {Fattorini}, {Ferrara}, {Foffano}, {Fonseca}, {Font}, {Fruck}, {Fukami}, {Garc{\'\i}a L{\'o}pez}, {Garczarczyk}, {Gasparyan}, {Gaug}, {Giglietto}, {Giordano}, {Gliwny}, {Godinovi{\'c}}, {Green}, {Green}, {Hadasch}, {Hahn}, {Heckmann}, {Herrera}, {Hoang},
  {Hrupec}, {H{\"u}tten}, {Inada}, {Inoue}, {Ishio}, {Iwamura}, {Jormanainen}, {Jouvin}, {Kajiwara}, {Karjalainen}, {Kerszberg}, {Kobayashi}, {Kubo}, {Kushida}, {Lamastra}, {Lelas}, {Leone}, {Lindfors}, {Lombardi}, {Longo}, {L{\'o}pez-Coto}, {L{\'o}pez-Moya}, {L{\'o}pez-Oramas}, {Loporchio}, {Machado de Oliveira Fraga}, {Maggio}, {Majumdar}, {Makariev}, {Mallamaci}, {Maneva}, {Manganaro}, {Mannheim}, {Maraschi}, {Mariotti}, {Mart{\'\i}nez}, {Mazin}, {Mender}, {Mi{\'c}anovi{\'c}}, {Miceli}, {Miener}, {Minev}, {Miranda}, {Mirzoyan}, {Molina}, {Moralejo}, {Morcuende}, {Moreno}, {Moretti}, {Neustroev}, {Nigro}, {Nilsson}, {Ninci}, {Nishijima}, {Noda}, {Nozaki}, {Ohtani}, {Oka}, {Otero-Santos}, {Paiano}, {Palatiello}, {Paneque}, {Paoletti}, {Paredes}, {Pavleti{\'c}}, {Pe{\~n}il}, {Perennes}, {Persic}, {Prada Moroni}, {Prandini}, {Priyadarshi}, {Puljak}, {Rhode}, {Rib{\'o}}, {Rico}, {Righi}, {Rugliancich}, {Saha}, {Sahakyan}, {Saito}, {Sakurai}, {Satalecka}, {Saturni}, {Schleicher}, {Schmidt}, {Schweizer},
  {Sitarek}, {{\v{S}}nidari{\'c}}, {Sobczynska}, {Spolon}, {Stamerra}, {Strom}, {Strzys}, {Suda}, {Suri{\'c}}, {Takahashi}, {Tavecchio}, {Temnikov}, {Terzi{\'c}}, {Teshima}, {Torres-Alb{\`a}}, {Tosti}, {Truzzi}, {Tutone}, {van Scherpenberg}, {Vanzo}, {Vazquez Acosta}, {Ventura}, {Verguilov}, {Vigorito}, {Vitale}, {Vovk}, {Will}, {Zari{\'c}}, {Angioni}, {D'Ammando}, {Ciprini}, {Cheung}, {Orienti}, {Pacciani}, {Prajapati}, {Kumar}, {Ganesh}, {Minev}, {Kurtenkov}, {Marchini}, {Carrasco}, {Escobedo}, {Porras}, {Recillas}, {L{\"a}hteenm{\"a}ki}, {Tornikoski}, {Berton}, {Tammi}, {Vera}, {Jorstad}, {Marscher}, {Weaver}, {Hart}, {Hallum}, {Larionov}, {Borman}, {Grishina}, {Kopatskaya}, {Larionova}, {Nikiforova}, {Morozova}, {Savchenko}, {Troitskaya}, {Troitsky}, {Vasilyev}, {Hodges}, {Hovatta}, {Kiehlmann}, {Max-Moerbeck}, {Readhead}, {Reeves}, \& {Pearson}}]{2021A&A...647A.163M}
{MAGIC Collaboration}, {Acciari}, V.~A., {Ansoldi}, S., {et~al.} 2021, \aap, 647, A163

\bibitem[{{Mankuzhiyil} {et~al.}(2011){Mankuzhiyil}, {Ansoldi}, {Persic}, \& {Tavecchio}}]{2011ApJ...733...14M}
{Mankuzhiyil}, N., {Ansoldi}, S., {Persic}, M., \& {Tavecchio}, F. 2011, \apj, 733, 14

\bibitem[{{Maraschi} {et~al.}(1992){Maraschi}, {Ghisellini}, \& {Celotti}}]{Maraschi92}
{Maraschi}, L., {Ghisellini}, G., \& {Celotti}, A. 1992, \apjl, 397, L5

\bibitem[{{Marin} {et~al.}(2013){Marin}, {Porquet}, {Goosmann}, {Dov{\v{c}}iak}, {Muleri}, {Grosso}, \& {Karas}}]{2013MNRAS.436.1615M}
{Marin}, F., {Porquet}, D., {Goosmann}, R.~W., {et~al.} 2013, \mnras, 436, 1615

\bibitem[{{Marscher} \& {Gear}(1985)}]{Marscher85}
{Marscher}, A.~P. \& {Gear}, W.~K. 1985, \apj, 298, 114

\bibitem[{{Massaro} {et~al.}(2004{\natexlab{a}}){Massaro}, {Perri}, {Giommi}, \& {Nesci}}]{2004A&A...413..489M}
{Massaro}, E., {Perri}, M., {Giommi}, P., \& {Nesci}, R. 2004{\natexlab{a}}, \aap, 413, 489

\bibitem[{{Massaro} {et~al.}(2004{\natexlab{b}}){Massaro}, {Perri}, {Giommi}, {Nesci}, \& {Verrecchia}}]{2004A&A...422..103M}
{Massaro}, E., {Perri}, M., {Giommi}, P., {Nesci}, R., \& {Verrecchia}, F. 2004{\natexlab{b}}, \aap, 422, 103

\bibitem[{{Matt} {et~al.}(2003){Matt}, {Guainazzi}, \& {Maiolino}}]{2003MNRAS.342..422M}
{Matt}, G., {Guainazzi}, M., \& {Maiolino}, R. 2003, \mnras, 342, 422

\bibitem[{{Mattox} {et~al.}(1996){Mattox}, {Bertsch}, {Chiang}, {Dingus}, {Digel}, {Esposito}, {Fierro}, {Hartman}, {Hunter}, {Kanbach}, {Kniffen}, {Lin}, {Macomb}, {Mayer-Hasselwander}, {Michelson}, {von Montigny}, {Mukherjee}, {Nolan}, {Ramanamurthy}, {Schneid}, {Sreekumar}, {Thompson}, \& {Willis}}]{1996ApJ...461..396M}
{Mattox}, J.~R., {Bertsch}, D.~L., {Chiang}, J., {et~al.} 1996, \apj, 461, 396

\bibitem[{{McElroy} {et~al.}(2016){McElroy}, {Husemann}, {Croom}, {Davis}, {Bennert}, {Busch}, {Combes}, {Eckart}, {Perez-Torres}, {Powell}, {Scharw{\"a}chter}, {Tremblay}, \& {Urrutia}}]{2016A&A...593L...8M}
{McElroy}, R.~E., {Husemann}, B., {Croom}, S.~M., {et~al.} 2016, \aap, 593, L8

\bibitem[{{Mishra} {et~al.}(2021){Mishra}, {Dai}, {Chen}, {Cheng}, {Jayasinghe}, {Tucker}, {Vallely}, {Bersier}, {Bose}, {Do}, {Dong}, {Holoien}, {Huber}, {Kochanek}, {Liang}, {Payne}, {Prieto}, {Shappee}, {Stanek}, {Bhatiani}, {Cox}, {DeFrancesco}, {Shen}, {Thompson}, \& {Wang}}]{2021ApJ...913..146M}
{Mishra}, H.~D., {Dai}, X., {Chen}, P., {et~al.} 2021, \apj, 913, 146

\bibitem[{{Narayan} \& {Yi}(1995)}]{1995ApJ...444..231N}
{Narayan}, R. \& {Yi}, I. 1995, \apj, 444, 231

\bibitem[{{Ouyang} {et~al.}(2021){Ouyang}, {Xiao}, {Zheng}, {Xu}, \& {Fan}}]{2021Ap&SS.366...12O}
{Ouyang}, Z., {Xiao}, H., {Zheng}, Y., {Xu}, P., \& {Fan}, J. 2021, \apss, 366, 12

\bibitem[{{Pandey} {et~al.}(2023){Pandey}, {Kushwaha}, {Wiita}, {Prince}, {Czerny}, \& {Stalin}}]{2023arXiv231005096P}
{Pandey}, A., {Kushwaha}, P., {Wiita}, P.~J., {et~al.} 2023, arXiv e-prints, arXiv:2310.05096

\bibitem[{{Parker} {et~al.}(2016){Parker}, {Komossa}, {Kollatschny}, {Walton}, {Schartel}, {Santos-Lle{\'o}}, {Harrison}, {Fabian}, {Zetzl}, {Grupe}, {Rodr{\'\i}guez-Pascual}, \& {Vasudevan}}]{2016MNRAS.461.1927P}
{Parker}, M.~L., {Komossa}, S., {Kollatschny}, W., {et~al.} 2016, \mnras, 461, 1927

\bibitem[{{Peterson}(1997)}]{1997iagn.book.....P}
{Peterson}, B.~M. 1997, {An Introduction to Active Galactic Nuclei}

\bibitem[{{Planck Collaboration} {et~al.}(2016){Planck Collaboration}, {Ade}, {Aghanim}, {Arnaud}, {Ashdown}, {Aumont}, {Baccigalupi}, {Banday}, {Barreiro}, {Bartlett}, {Bartolo}, {Battaner}, {Battye}, {Benabed}, {Beno{\^\i}t}, {Benoit-L{\'e}vy}, {Bernard}, {Bersanelli}, {Bielewicz}, {Bock}, {Bonaldi}, {Bonavera}, {Bond}, {Borrill}, {Bouchet}, {Boulanger}, {Bucher}, {Burigana}, {Butler}, {Calabrese}, {Cardoso}, {Catalano}, {Challinor}, {Chamballu}, {Chary}, {Chiang}, {Chluba}, {Christensen}, {Church}, {Clements}, {Colombi}, {Colombo}, {Combet}, {Coulais}, {Crill}, {Curto}, {Cuttaia}, {Danese}, {Davies}, {Davis}, {de Bernardis}, {de Rosa}, {de Zotti}, {Delabrouille}, {D{\'e}sert}, {Di Valentino}, {Dickinson}, {Diego}, {Dolag}, {Dole}, {Donzelli}, {Dor{\'e}}, {Douspis}, {Ducout}, {Dunkley}, {Dupac}, {Efstathiou}, {Elsner}, {En{\ss}lin}, {Eriksen}, {Farhang}, {Fergusson}, {Finelli}, {Forni}, {Frailis}, {Fraisse}, {Franceschi}, {Frejsel}, {Galeotta}, {Galli}, {Ganga}, {Gauthier}, {Gerbino}, {Ghosh}, {Giard},
  {Giraud-H{\'e}raud}, {Giusarma}, {Gjerl{\o}w}, {Gonz{\'a}lez-Nuevo}, {G{\'o}rski}, {Gratton}, {Gregorio}, {Gruppuso}, {Gudmundsson}, {Hamann}, {Hansen}, {Hanson}, {Harrison}, {Helou}, {Henrot-Versill{\'e}}, {Hern{\'a}ndez-Monteagudo}, {Herranz}, {Hildebrandt}, {Hivon}, {Hobson}, {Holmes}, {Hornstrup}, {Hovest}, {Huang}, {Huffenberger}, {Hurier}, {Jaffe}, {Jaffe}, {Jones}, {Juvela}, {Keih{\"a}nen}, {Keskitalo}, {Kisner}, {Kneissl}, {Knoche}, {Knox}, {Kunz}, {Kurki-Suonio}, {Lagache}, {L{\"a}hteenm{\"a}ki}, {Lamarre}, {Lasenby}, {Lattanzi}, {Lawrence}, {Leahy}, {Leonardi}, {Lesgourgues}, {Levrier}, {Lewis}, {Liguori}, {Lilje}, {Linden-V{\o}rnle}, {L{\'o}pez-Caniego}, {Lubin}, {Mac{\'\i}as-P{\'e}rez}, {Maggio}, {Maino}, {Mandolesi}, {Mangilli}, {Marchini}, {Maris}, {Martin}, {Martinelli}, {Mart{\'\i}nez-Gonz{\'a}lez}, {Masi}, {Matarrese}, {McGehee}, {Meinhold}, {Melchiorri}, {Melin}, {Mendes}, {Mennella}, {Migliaccio}, {Millea}, {Mitra}, {Miville-Desch{\^e}nes}, {Moneti}, {Montier}, {Morgante}, {Mortlock},
  {Moss}, {Munshi}, {Murphy}, {Naselsky}, {Nati}, {Natoli}, {Netterfield}, {N{\o}rgaard-Nielsen}, {Noviello}, {Novikov}, {Novikov}, {Oxborrow}, {Paci}, {Pagano}, {Pajot}, {Paladini}, {Paoletti}, {Partridge}, {Pasian}, {Patanchon}, {Pearson}, {Perdereau}, {Perotto}, {Perrotta}, {Pettorino}, {Piacentini}, {Piat}, {Pierpaoli}, {Pietrobon}, {Plaszczynski}, {Pointecouteau}, {Polenta}, {Popa}, {Pratt}, {Pr{\'e}zeau}, {Prunet}, {Puget}, {Rachen}, {Reach}, {Rebolo}, {Reinecke}, {Remazeilles}, {Renault}, {Renzi}, {Ristorcelli}, {Rocha}, {Rosset}, {Rossetti}, {Roudier}, {Rouill{\'e} d'Orfeuil}, {Rowan-Robinson}, {Rubi{\~n}o-Mart{\'\i}n}, {Rusholme}, {Said}, {Salvatelli}, {Salvati}, {Sandri}, {Santos}, {Savelainen}, {Savini}, {Scott}, {Seiffert}, {Serra}, {Shellard}, {Spencer}, {Spinelli}, {Stolyarov}, {Stompor}, {Sudiwala}, {Sunyaev}, {Sutton}, {Suur-Uski}, {Sygnet}, {Tauber}, {Terenzi}, {Toffolatti}, {Tomasi}, {Tristram}, {Trombetti}, {Tucci}, {Tuovinen}, {T{\"u}rler}, {Umana}, {Valenziano}, {Valiviita}, {Van Tent},
  {Vielva}, {Villa}, {Wade}, {Wandelt}, {Wehus}, {White}, {White}, {Wilkinson}, {Yvon}, {Zacchei}, \& {Zonca}}]{2016A&A...594A..13P}
{Planck Collaboration}, {Ade}, P.~A.~R., {Aghanim}, N., {et~al.} 2016, \aap, 594, A13

\bibitem[{{P{\"o}ssel}(2020)}]{2020OJAp....3E...2P}
{P{\"o}ssel}, M. 2020, The Open Journal of Astrophysics, 3, 2

\bibitem[{{Prandini} \& {Ghisellini}(2022)}]{2022Galax..10...35P}
{Prandini}, E. \& {Ghisellini}, G. 2022, Galaxies, 10, 35

\bibitem[{{Prince} {et~al.}(2021){Prince}, {Khatoon}, \& {Stalin}}]{2021MNRAS.502.5245P}
{Prince}, R., {Khatoon}, R., \& {Stalin}, C.~S. 2021, \mnras, 502, 5245

\bibitem[{{Prince} {et~al.}(2018){Prince}, {Raman}, {Hahn}, {Gupta}, \& {Majumdar}}]{2018ApJ...866...16P}
{Prince}, R., {Raman}, G., {Hahn}, J., {Gupta}, N., \& {Majumdar}, P. 2018, \apj, 866, 16

\bibitem[{{Raimundo} {et~al.}(2019){Raimundo}, {Vestergaard}, {Koay}, {Lawther}, {Casasola}, \& {Peterson}}]{2019MNRAS.486..123R}
{Raimundo}, S.~I., {Vestergaard}, M., {Koay}, J.~Y., {et~al.} 2019, \mnras, 486, 123

\bibitem[{{Raiteri} {et~al.}(2015){Raiteri}, {Stamerra}, {Villata}, {Larionov}, {Acosta-Pulido}, {Ar{\'e}valo}, {Arkharov}, {Bachev}, {Ben{\'\i}tez}, {Bozhilov}, {Borman}, {Buemi}, {Calcidese}, {Carnerero}, {Carosati}, {Chigladze}, {Damljanovic}, {Di Paola}, {Doroshenko}, {Efimova}, {Ehgamberdiev}, {Giroletti}, {Gonz{\'a}lez-Morales}, {Grinon-Marin}, {Grishina}, {Hiriart}, {Ibryamov}, {Klimanov}, {Kopatskaya}, {Kurtanidze}, {Kurtanidze}, {Kurtenkov}, {Larionova}, {Larionova}, {L{\'a}zaro}, {L{\"a}hteenm{\"a}ki}, {Leto}, {Markovic}, {Mirzaqulov}, {Mokrushina}, {Morozova}, {M{\'u}jica}, {Nazarov}, {Nikolashvili}, {Ohlert}, {Ovcharov}, {Paiano}, {Pastor Yabar}, {Prandini}, {Ramakrishnan}, {Sadun}, {Semkov}, {Sigua}, {Strigachev}, {Tammi}, {Tornikoski}, {Trigilio}, {Troitskaya}, {Troitsky}, {Umana}, {Velasco}, \& {Vince}}]{2015MNRAS.454..353R}
{Raiteri}, C.~M., {Stamerra}, A., {Villata}, M., {et~al.} 2015, \mnras, 454, 353

\bibitem[{{Ricci} {et~al.}(2016){Ricci}, {Bauer}, {Arevalo}, {Boggs}, {Brandt}, {Christensen}, {Craig}, {Gandhi}, {Hailey}, {Harrison}, {Koss}, {Markwardt}, {Stern}, {Treister}, \& {Zhang}}]{2016ApJ...820....5R}
{Ricci}, C., {Bauer}, F.~E., {Arevalo}, P., {et~al.} 2016, \apj, 820, 5

\bibitem[{{Ricci} \& {Trakhtenbrot}(2023)}]{2023NatAs...7.1282R}
{Ricci}, C. \& {Trakhtenbrot}, B. 2023, Nature Astronomy, 7, 1282

\bibitem[{{Rivers} {et~al.}(2015){Rivers}, {Balokovi{\'c}}, {Ar{\'e}valo}, {Bauer}, {Boggs}, {Brandt}, {Brightman}, {Christensen}, {Craig}, {Gandhi}, {Hailey}, {Harrison}, {Koss}, {Ricci}, {Stern}, {Walton}, \& {Zhang}}]{2015ApJ...815...55R}
{Rivers}, E., {Balokovi{\'c}}, M., {Ar{\'e}valo}, P., {et~al.} 2015, \apj, 815, 55

\bibitem[{{Rodr{\'\i}guez-Pascual} {et~al.}(1997){Rodr{\'\i}guez-Pascual}, {Alloin}, {Clavel}, {Crenshaw}, {Horne}, {Kriss}, {Krolik}, {Malkan}, {Netzer}, {O'Brien}, {Peterson}, {Reichert}, {Wamsteker}, {Alexander}, {Barr}, {Blandford}, {Bregman}, {Carone}, {Clements}, {Courvoisier}, {De Robertis}, {Dietrich}, {Dottori}, {Edelson}, {Filippenko}, {Gaskell}, {Huchra}, {Hutchings}, {Kollatschny}, {Koratkar}, {Korista}, {Laor}, {MacAlpine}, {Martin}, {Maoz}, {McCollum}, {Morris}, {Perola}, {Pogge}, {Ptak}, {Recondo-Gonz{\'a}lez}, {Rodr{\'\i}guez-Espinoza}, {Rokaki}, {Santos-Lle{\'o}}, {Sekiguchi}, {Shull}, {Snijders}, {Sparke}, {Stirpe}, {Stoner}, {Sun}, {Wagner}, {Wanders}, {Wilkes}, {Winge}, \& {Zheng}}]{1997ApJS..110....9R}
{Rodr{\'\i}guez-Pascual}, P.~M., {Alloin}, D., {Clavel}, J., {et~al.} 1997, \apjs, 110, 9

\bibitem[{{Roming} {et~al.}(2005){Roming}, {Kennedy}, {Mason}, {Nousek}, {Ahr}, {Bingham}, {Broos}, {Carter}, {Hancock}, {Huckle}, {Hunsberger}, {Kawakami}, {Killough}, {Koch}, {McLelland}, {Smith}, {Smith}, {Soto}, {Boyd}, {Breeveld}, {Holland}, {Ivanushkina}, {Pryzby}, {Still}, \& {Stock}}]{2005SSRv..120...95R}
{Roming}, P. W.~A., {Kennedy}, T.~E., {Mason}, K.~O., {et~al.} 2005, \ssr, 120, 95

\bibitem[{{Ruan} {et~al.}(2016){Ruan}, {Anderson}, {Cales}, {Eracleous}, {Green}, {Morganson}, {Runnoe}, {Shen}, {Wilkinson}, {Blanton}, {Dwelly}, {Georgakakis}, {Greene}, {LaMassa}, {Merloni}, \& {Schneider}}]{2016ApJ...826..188R}
{Ruan}, J.~J., {Anderson}, S.~F., {Cales}, S.~L., {et~al.} 2016, \apj, 826, 188

\bibitem[{{Runnoe} {et~al.}(2016){Runnoe}, {Cales}, {Ruan}, {Eracleous}, {Anderson}, {Shen}, {Green}, {Morganson}, {LaMassa}, {Greene}, {Dwelly}, {Schneider}, {Merloni}, {Georgakakis}, \& {Roman-Lopes}}]{2016MNRAS.455.1691R}
{Runnoe}, J.~C., {Cales}, S., {Ruan}, J.~J., {et~al.} 2016, \mnras, 455, 1691

\bibitem[{{Sambruna} {et~al.}(1999){Sambruna}, {Ghisellini}, {Hooper}, {Kollgaard}, {Pesce}, \& {Urry}}]{1999ApJ...515..140S}
{Sambruna}, R.~M., {Ghisellini}, G., {Hooper}, E., {et~al.} 1999, \apj, 515, 140

\bibitem[{{Sbarrato} {et~al.}(2012){Sbarrato}, {Ghisellini}, {Maraschi}, \& {Colpi}}]{2012MNRAS.421.1764S}
{Sbarrato}, T., {Ghisellini}, G., {Maraschi}, L., \& {Colpi}, M. 2012, \mnras, 421, 1764

\bibitem[{{Shakura} \& {Sunyaev}(1973)}]{1973A&A....24..337S}
{Shakura}, N.~I. \& {Sunyaev}, R.~A. 1973, \aap, 24, 337

\bibitem[{{Shappee} {et~al.}(2014){Shappee}, {Prieto}, {Grupe}, {Kochanek}, {Stanek}, {De Rosa}, {Mathur}, {Zu}, {Peterson}, {Pogge}, {Komossa}, {Im}, {Jencson}, {Holoien}, {Basu}, {Beacom}, {Szczygie{\l}}, {Brimacombe}, {Adams}, {Campillay}, {Choi}, {Contreras}, {Dietrich}, {Dubberley}, {Elphick}, {Foale}, {Giustini}, {Gonzalez}, {Hawkins}, {Howell}, {Hsiao}, {Koss}, {Leighly}, {Morrell}, {Mudd}, {Mullins}, {Nugent}, {Parrent}, {Phillips}, {Pojmanski}, {Rosing}, {Ross}, {Sand}, {Terndrup}, {Valenti}, {Walker}, \& {Yoon}}]{2014ApJ...788...48S}
{Shappee}, B.~J., {Prieto}, J.~L., {Grupe}, D., {et~al.} 2014, \apj, 788, 48

\bibitem[{{Sikora} {et~al.}(1994){Sikora}, {Begelman}, \& {Rees}}]{Sikora94}
{Sikora}, M., {Begelman}, M.~C., \& {Rees}, M.~J. 1994, \apj, 421, 153

\bibitem[{{Sniegowska} {et~al.}(2020){Sniegowska}, {Czerny}, {Bon}, \& {Bon}}]{2020A&A...641A.167S}
{Sniegowska}, M., {Czerny}, B., {Bon}, E., \& {Bon}, N. 2020, \aap, 641, A167

\bibitem[{{Stocke} {et~al.}(2011){Stocke}, {Danforth}, \& {Perlman}}]{2011ApJ...732..113S}
{Stocke}, J.~T., {Danforth}, C.~W., \& {Perlman}, E.~S. 2011, \apj, 732, 113

\bibitem[{{Stocke} {et~al.}(1991){Stocke}, {Morris}, {Gioia}, {Maccacaro}, {Schild}, {Wolter}, {Fleming}, \& {Henry}}]{1991ApJS...76..813S}
{Stocke}, J.~T., {Morris}, S.~L., {Gioia}, I.~M., {et~al.} 1991, \apjs, 76, 813

\bibitem[{{Storchi-Bergmann} {et~al.}(1993){Storchi-Bergmann}, {Baldwin}, \& {Wilson}}]{1993ApJ...410L..11S}
{Storchi-Bergmann}, T., {Baldwin}, J.~A., \& {Wilson}, A.~S. 1993, \apjl, 410, L11

\bibitem[{{Tramacere} {et~al.}(2009){Tramacere}, {Giommi}, {Perri}, {Verrecchia}, \& {Tosti}}]{2009A&A...501..879T}
{Tramacere}, A., {Giommi}, P., {Perri}, M., {Verrecchia}, F., \& {Tosti}, G. 2009, \aap, 501, 879

\bibitem[{{Tramacere} {et~al.}(2011){Tramacere}, {Massaro}, \& {Taylor}}]{2011ApJ...739...66T}
{Tramacere}, A., {Massaro}, E., \& {Taylor}, A.~M. 2011, \apj, 739, 66

\bibitem[{{Tran} {et~al.}(1992){Tran}, {Osterbrock}, \& {Martel}}]{1992AJ....104.2072T}
{Tran}, H.~D., {Osterbrock}, D.~E., \& {Martel}, A. 1992, \aj, 104, 2072

\bibitem[{{Turner} {et~al.}(2018){Turner}, {Reeves}, {Braito}, {Lobban}, {Kraemer}, \& {Miller}}]{2018MNRAS.481.2470T}
{Turner}, T.~J., {Reeves}, J.~N., {Braito}, V., {et~al.} 2018, \mnras, 481, 2470

\bibitem[{{Urry}(1998)}]{Urry98}
{Urry}, C.~M. 1998, Advances in Space Research, 21, 89

\bibitem[{{Urry} \& {Padovani}(1995)}]{1995PASP..107..803U}
{Urry}, C.~M. \& {Padovani}, P. 1995, \pasp, 107, 803

\bibitem[{{Vaughan} {et~al.}(2003){Vaughan}, {Edelson}, {Warwick}, \& {Uttley}}]{2003MNRAS.345.1271V}
{Vaughan}, S., {Edelson}, R., {Warwick}, R.~S., \& {Uttley}, P. 2003, \mnras, 345, 1271

\bibitem[{{Wagner} \& {Witzel}(1995)}]{1995ARA&A..33..163W}
{Wagner}, S.~J. \& {Witzel}, A. 1995, \araa, 33, 163

\bibitem[{{Wang} {et~al.}(2020){Wang}, {Xu}, \& {Wei}}]{2020ApJ...901....1W}
{Wang}, J., {Xu}, D.~W., \& {Wei}, J.~Y. 2020, \apj, 901, 1

\bibitem[{{Wang} {et~al.}(2002){Wang}, {Staubert}, \& {Ho}}]{2002ApJ...579..554W}
{Wang}, J.-M., {Staubert}, R., \& {Ho}, L.~C. 2002, \apj, 579, 554

\bibitem[{{Wood} {et~al.}(2017){Wood}, {Caputo}, {Charles}, {Di Mauro}, {Magill}, {Perkins}, \& {Fermi-LAT Collaboration}}]{2017ICRC...35..824W}
{Wood}, M., {Caputo}, R., {Charles}, E., {et~al.} 2017, in International Cosmic Ray Conference, Vol. 301, 35th International Cosmic Ray Conference (ICRC2017), 824

\bibitem[{{Wright}(2001)}]{2001ApJ...553..538W}
{Wright}, E.~L. 2001, \apj, 553, 538

\bibitem[{{Xiao} {et~al.}(2022){Xiao}, {Fan}, {Ouyang}, {Hu}, {Chen}, {Fu}, \& {Zhang}}]{2022ApJ...936..146X}
{Xiao}, H., {Fan}, J., {Ouyang}, Z., {et~al.} 2022, \apj, 936, 146

\bibitem[{{Xiong} \& {Zhang}(2014)}]{2014MNRAS.441.3375X}
{Xiong}, D.~R. \& {Zhang}, X. 2014, \mnras, 441, 3375

\bibitem[{{Yan} {et~al.}(2012){Yan}, {Zeng}, \& {Zhang}}]{2012PASJ...64...80Y}
{Yan}, D., {Zeng}, H., \& {Zhang}, L. 2012, \pasj, 64, 80

\bibitem[{{Yan} {et~al.}(2014){Yan}, {Zeng}, \& {Zhang}}]{2014MNRAS.439.2933Y}
{Yan}, D., {Zeng}, H., \& {Zhang}, L. 2014, \mnras, 439, 2933

\bibitem[{{Yang} {et~al.}(2018){Yang}, {Wu}, {Fan}, {Jiang}, {McGreer}, {Shangguan}, {Yao}, {Wang}, {Joshi}, {Green}, {Wang}, {Feng}, {Fu}, {Yang}, \& {Liu}}]{2018ApJ...862..109Y}
{Yang}, Q., {Wu}, X.-B., {Fan}, X., {et~al.} 2018, \apj, 862, 109

\bibitem[{{Yuan} \& {Narayan}(2014)}]{2014ARA&A..52..529Y}
{Yuan}, F. \& {Narayan}, R. 2014, \araa, 52, 529

\bibitem[{{Zeng} {et~al.}(2022){Zeng}, {Yan}, {Hu}, \& {Wang}}]{2022MNRAS.511..938Z}
{Zeng}, Y., {Yan}, D., {Hu}, W., \& {Wang}, J. 2022, \mnras, 511, 938

\bibitem[{{Zhang} {et~al.}(2012){Zhang}, {Liang}, {Zhang}, \& {Bai}}]{2012ApJ...752..157Z}
{Zhang}, J., {Liang}, E.-W., {Zhang}, S.-N., \& {Bai}, J.~M. 2012, \apj, 752, 157

\bibitem[{{Zhang} {et~al.}(2014){Zhang}, {Sun}, {Liang}, {Lu}, {Lu}, \& {Zhang}}]{2014ApJ...788..104Z}
{Zhang}, J., {Sun}, X.-N., {Liang}, E.-W., {et~al.} 2014, \apj, 788, 104

\bibitem[{{Zheng} {et~al.}(2019){Zheng}, {Kang}, {Yang}, \& {Bai}}]{2019ApJ...873....7Z}
{Zheng}, Y.~G., {Kang}, S.~J., {Yang}, C.~Y., \& {Bai}, J.~M. 2019, \apj, 873, 7

\bibitem[{{Zheng} {et~al.}(2020){Zheng}, {Kang}, {Yang}, \& {Bai}}]{2020MNRAS.499.1188Z}
{Zheng}, Y.~G., {Kang}, S.~J., {Yang}, C.~Y., \& {Bai}, J.~M. 2020, \mnras, 499, 1188

\bibitem[{{Zheng} {et~al.}(2017){Zheng}, {Yang}, {Zhang}, \& {Wang}}]{2017ApJS..228....1Z}
{Zheng}, Y.~G., {Yang}, C.~Y., {Zhang}, L., \& {Wang}, J.~C. 2017, \apjs, 228, 1

\bibitem[{{Zhou} {et~al.}(2021){Zhou}, {Zheng}, {Zhu}, \& {Kang}}]{2021ApJ...915...59Z}
{Zhou}, R.~X., {Zheng}, Y.~G., {Zhu}, K.~R., \& {Kang}, S.~J. 2021, \apj, 915, 59

\bibitem[{{Zhu} {et~al.}(2021){Zhu}, {Kang}, {Zhou}, \& {Zheng}}]{2021ApJ...916...93Z}
{Zhu}, K.~R., {Kang}, S.~J., {Zhou}, R.~X., \& {Zheng}, Y.~G. 2021, \apj, 916, 93

\end{thebibliography}

\end{document}